\documentclass[12pt]{article}
\pdfoutput=1
\usepackage{jheppub}
\usepackage{amsfonts}
\usepackage{amsmath,amssymb} 
\usepackage{graphicx} 
\usepackage[dvipsnames,x11names]{xcolor} 

\usepackage[nottoc,notlof,notlot]{tocbibind} 
\usepackage[titles]{tocloft} 

\usepackage{float}
\usepackage{subcaption}
\usepackage{tikz}
\numberwithin{equation}{section} 

\def\Tr{\mbox{Tr}}

\title{\textbf{Quantum Quenches and Thermalization in SYK models}}

\author{Ritabrata
  Bhattacharya,}\emailAdd{ritabratabhattacharya@hri.res.in}

\author{Dileep P. Jatkar,}\emailAdd{dileep@hri.res.in}

\author{and Nilakash Sorokhaibam}\emailAdd{nilakashsingh@hri.res.in}

\affiliation{
  Harish-Chandra Research Institute, 
  Homi Bhabha National Institute\\
    Chhatnag Road, Jhunsi, 
    Allahabad 211019, India}

  \abstract{We study non-equilibrium dynamics in SYK models using
    quantum quench.  We consider models with two, four, and higher
    fermion interactions ($q=2, 4$, and higher) and use two different
    types of quench protocol, which we call step and bump quenches. We
    analyse evolution of fermion two-point functions without long time
    averaging.  We observe that in $q=2$ theory the two-point
    functions do not thermalize.  We find thermalization in $q=4$ and
    higher theories without long time averaging.  We calculate two
    different exponents of which one is equal to the coupling and the
    other is proportional to the final temperature.  This result is
    more robust than thermalization obtained from long time averaging
    as proposed by the eigenstate thermalization hypothesis(ETH).
    Thermalization achieved without long time averaging is more akin
    to mixing than ergodicity.}

\begin{document}
\maketitle

\section{Introduction and Summary}

The study of non-equilibrium dynamics is becoming important both in
condensed matter physics \cite{Gogolin:2016hwy, Traschen:1990sw,
  Cramer:PhysRevLett.100.030602, Kitaev:2015,Eberlein:2017wah,
  Erdmenger:2016msd,Cardy:2014rqa,Polkovnikov:2010yn,
  Calabrese:2007rg} as well as in string theory\cite{Mandal:2015jla,
  Das:2015jka,Goel:2018ubv,Hunter-Jones:2017raw,Mandal:2015kxi}.
One of the most interesting question in this field is to understand
patterns of thermalization in the systems which are out of
equilibrium.  For example, it is important to know under what
conditions a closed quantum system thermalizes, i.e., for a system
prepared in a pure excited state, and undergoes unitary evolution,
determine how the late time limit of the expectation values of certain
observables are effectively described by a thermal
ensemble\footnote{The expectation values can equilibrate but the
  stationary limits may not be described by a thermal ensemble, which
  we will observe below for $q=2$ theory for which the fermion
  two-point functions freeze instantaneously but its values are not
  described by a thermal ensemble.}.  Interest in the non-equilibrium
dynamics from string theory point-of-view stems from black hole
physics.  The AdS/CFT correspondence(or the holographic principle, in
general) says that a black hole corresponds to thermal ensemble in the
\emph{boundary} quantum theory, and the thermalization process in
the quantum system is conjectured to be dual to black hole formation
in the bulk gravitation theory.
 
On the bulk gravity side it has been conjectured that black holes are
fast scramblers \cite{Sekino:2008he}. This proposal led to another
conjecture\cite{Maldacena:2015waa} that the chaotic behaviour, that
leads to scrambling, which is parametrized by the Lyapunov exponent
$\lambda_L$ has an upper bound, and that upper bound is saturated by
black holes.  This naturally gave additional impetus to the study of
non-equilibrium dynamics in systems which exhibit chaos, especially if
the Lyapunov exponent of the theory saturates the upper bound.

The eigenstate thermalization hypothesis (ETH) is an attempt to
explain how closed unitary quantum systems in pure excited states can
thermalize\cite{Deutsch:PhysRevA.43.2046,Srednicki:1993im}.
Thermalization with ETH crucially involves long time averaging of the
observables under consideration. It is, however, not clear what is the
precise relation between chaos and ETH.  In many studies of quantum
systems, thermalization is observed even without long-time
averaging\cite{Cramer:PhysRevLett.100.030602}.  Thermalization has
also been seen in the integrable systems without long time averaging.
The late time behaviour of integrable models is described by the
generalized Gibbs ensembles\cite{fioretto2010quantum, Das:2017sgp}.
These ensembles have fugacities turned on for several conserved
charges of the integrable system.  The integrable model, by
definition, is not chaotic on its own.

The Sachdev-Ye-Kitaev(SYK) model which is a (0+1) dimensional model of
Majorana fermions with all to all $q$-body random interactions.  The
$q=4$ and higher models were studied by Kitaev\cite{Kitaev:2015}, and
by Maldacena and Stanford\cite{Maldacena:2016hyu}.  They showed that
the out of time ordered four point correlators in these models
saturate the upper bound on the Lyapunov exponent, in addition, these
models also satisfy ETH. There has been a lot of work on this model,
its variants and their bulk duals \cite{Kitaev:2015, Witten:2016iux,
  Maldacena:2016hyu, Mandal:2017thl, Sonner:2017hxc, Haque:2017bts,
  Stanford:2017thb, Berkooz:2016cvq, Eberlein:2017wah,
  Murugan:2017eto, Klebanov:2017nlk, Kourkoulou:2017zaj,
  Erdmenger:2016msd, Narayan:2017qtw, Das:2017pif, Turiaci:2017zwd,
  Klebanov:2016xxf, Krishnan:2016bvg, Krishnan:2017lra,
  Callebaut:2018nlq, Goel:2018ubv, Gaikwad:2018dfc, Choudhury:2017tax,
  Das:2017wae, Bulycheva:2017ilt, Narayan:2017hvh, Klebanov:2018fzb,
  Kitaev:2017awl, Das:2017hrt, Engelsoy:2016xyb, Jensen:2016pah,
  Roychowdhury:2018clp}.  The $q=2$ SYK model is not chaotic and also
does not satisfy ETH. However, unlike integrable local quantum
systems, it does not have infinite number of conserved charges in
spite of them being exactly solvable. In fact, it has only one
conserved charge which is the total energy of the system. With this
background in mind, we study non-equilibrium dynamics of excited
states in $q=2, 4$, and higher SYK models.

The most convenient method for studying non-equilibrium dynamics, both
theoretically
\cite{Calabrese:2006rx,Calabrese:2007rg,Polkovnikov:2010yn,
  ziraldo2013thermalization,Cardy:2014rqa,Das:2015jka,
  Calabrese:2016xau,Eberlein:2017wah,Erdmenger:2016msd,
  Mandal:2015kxi} and
experimentally\cite{greiner2002collapse,kinoshita2006quantum}, turns
out to be quantum quench.  In other words, quantum quenches are the
most convenient way of generating non-trivial excited states of the
theory. In quantum quench one abruptly changes parameters of the
Hamiltonian of the system starting from an equilibrium
configuration(generally a thermal state or the ground state) of the
system.  The change in the coupling generally excites the system and
the system evolves non-trivially with the final Hamiltonian. The
evolution of the system is examined by calculating the expectation
values of some of the observables of the system. If the expectation
values of those observables approach the expectation values in a
thermal ensemble, the system is said to have thermalized.

Certain aspects of quantum quenches in SYK models have been studied in
\cite{Eberlein:2017wah}. In this paper, using similar numerical
techniques, we will study quantum quenches in $q=2, 4$, and higher SYK
models. We will consider one particular observable which is the
greater Green's function $G^>(t_1,t_2)$.
\begin{equation}
G^{>}(t_1,t_2)=-i\sum_{i=1}^N\langle\psi_i(t_2)\psi_i(t_1)\rangle \\
\end{equation}
For majorana fermions, all other two-point functions can be calculated
from $G^>(t_1,t_2)$. The non-trivial time evolution of $G^>(t_1,t_2)$
can be examined by exactly solving its equations of motion which are
the Kadanoff-Baym(KB) equations. Our analysis will involve changing
various parameters with two different kinds of time dependence. The
usual quench protocol in condensed-matter literature is changing,
suddenly\footnote{The smallest scale in the sudden limit is the time
  scale over which the couplings change.} or smoothly but rapidly,
the parameters from one value to another different value. We will
consider sudden change from one value to another, which we call step
quench. In addition to this, we will also study bump quench, in which
the coupling changes for a finite time interval before returning back
to the original value\footnote{Although bump quenches are not well
  studied in condensed-matter literature, they are more relevant to
  black hole physics (using AdS/CFT) than step quenches
  \cite{Bhattacharyya:2009uu, Caceres:2014pda}.}.  We follow the
convention $q=k$ quench when the final hamiltonian of the system has
$k$ fermion interaction and the couplings $J_q$ undergo quench with
$q\not= k$. We will also consider only sudden limit for both step and
bump quenches.

\vspace{0.5cm}

The quenches which are relevant for our main results are:
\begin{itemize}
\item Quenches in $q=2$ theory: We use four, six and eight fermion
  interactions ($J_4$, $J_6$, and $J_8$ couplings) separately to
  quench the system for both step and bump protocols.
\item Quenches in $q=4$ theory: For this theory, we use $J_2$ (two
  fermion interaction coupling), $J_6$, and $J_8$ with both step and
  bump protocols.
\end{itemize}

For quenches in $q=4$ theory, we start from finite temperature thermal
states which reduce finite-size effect drastically and ensure good
numerical accuracy. For quenches in $q=2$ theory, we start from the
ground state as well as from finite temperature states. We observe
that if we take the initial thermal states to be of sufficiently low
temperatures, the effect of the initial temperature becomes
insignificant. We consider only the greater Green's function
$G^>(t_1,t_2)$, because for the Majorana fermions, all other two point
functions can be expressed in term of $G^>(t_1,t_2)$ alone.

\vspace{0.5cm}

The main technical results of our analysis are as follows:
\begin{itemize}
\item In $q=2$ theory, the two point functions do not thermalize in
  all the quench scenario. But an interesting observation is that the
  two point functions equilibrate instantaneously as soon as both the
  time arguments are outside the quench region.
\item In $q=4$ theory, the two point functions thermalize
  for all the quench scenario.  $G^>(t_1,t_2)$ converges exponentially
  towards its equilibrium expectation value. This exponential
  behaviour is observed as soon as both the time arguments are outside
  the quench region.
\item In $q=4$ theory, we also identify two exponents, of which, one
  is equal to the coupling and the other is proportional to the final
  temperature. The first one is the exponent of $G^>(t-t_a,t)$ as a
  function of $t$ with $t_a$ fixed, while the other is the exponent of
  $G^>(t,t_b)$ as a function of $t$ with $t_b$ fixed.
\item We compute the thermalization rate of the effective temperature
  in both step and bump quench.  We show that the thermalization of the
  effective temperature fits an exponential ansatz. We also find that the thermalization rate is independent of the coupling constant.
\end{itemize}

An important aspect of the present work is to check if step quenches
produce special fine-tuned pure states which looks exactly
thermal. These pure states are inspired by the Euclidean evolved
boundary states of Calabrese and Cardy \cite{Calabrese:2005in}. These
states, which we will refer to as Kourkoulou-Maldacena (KM) states below,
have interesting bulk duals \cite{Kourkoulou:2017zaj}. The details of
these pure states can be found in section \ref{sec:calabr-cardy-stat}.
We observed that the final states of quantum quenches using disordered
couplings are not KM states. But one can use mass like terms to
perform the sudden step quenches for which the final states are the KM
states.

The thermalization we observe in $q=4$ theory without long time
averaging, is much more robust than what one expects from the ETH.  We
therefore believe that \emph{thermalization in a chaotic system is more
  akin to mixing in classical systems which is a stronger condition
  than ergodicity.}

The outline of this paper is as follows: In section
\ref{sec:quantum-quench-syk}, we will briefly recall the SYK model.
This will also be used to fix our notation. We will write down the
Schwinger-Dyson equation for a model with both $q=2$ and $4$
interactions. The couplings for these terms will have arbitrary time
dependence to start with.  We will then set up the Kadanoff-Baym
equations for this system which can be easily generalized for higher
$q$ models. Finally we will briefly discuss the eigenstate
thermalization hypothesis(ETH).  In section
\ref{sec:calabr-cardy-stat}, we discuss Kourkoulou-Maldacena states
with an eye on possible relation between our results and these excited
states.  In section \ref{sec:quantum-quench-model}, we discuss various
quench protocols that we study in the SYK model and present results of
our numerical computations.  Section \ref{sec:condis} contains
conclusion and discussion where we wrap up our results and discuss
about ways to prepare Kourkoulou-Maldacena states and the implications
of thermalization without long-time averaging.

\section{The SYK models}
\label{sec:quantum-quench-syk}

We begin with a review the model studied by Sachdev et
al.\cite{Eberlein:2017wah}.  This will help set up notation for
subsequent sections.  Our starting point is the SYK model with the
hamiltonian $H$ that contains $q$-point ($q$ even) interaction between
$N$ Majorana fermions,
\begin{equation}\label{eq:qq1}
H=(i)^{q/2}\sum_{1\leq i_1< i_2<...< i_q\leq
  N}J_{i_1,i_2,..,i_q}\psi_{i_1}\psi_{i_2}....\psi_{i_q}\
\end{equation}
The coupling $J_{i_1,i_2,..,i_q}$ is random with gaussian
distribution, with vanishing mean value and the width of the gaussian
is given by
\begin{equation}
  \label{eq:qq2}
  \langle J_{i_1,i_2,..,i_q}^2\rangle=\frac{J^2(q-1)!}{N^{q-1}}\ .
\end{equation}
To compute correlators at finite temperature the Schwinger-Keldysh
formalism is employed in which, the observables are computed by
integrating along the closed-time contour $\mathcal{C}$.  The initial
state is evolved along this contour both forward and backwards in
time.  The contour-ordered Green’s function is defined as
\cite{Eberlein:2017wah},
\begin{eqnarray}
iG(t_1,t_2)&=&\langle T_{\mathcal{C}}\left(
               \psi_{i}(t_1)\psi_i(t_2)\right)\rangle\nonumber\\ 
&=&\theta_{\mathcal{C}}(t_1-t_2)\langle
    \psi_{i}(t_1)\psi_i(t_2)\rangle -
    \theta_{\mathcal{C}}(t_2-t_1)\langle\psi_{i}(t_2)\psi_i(t_1)\rangle\ 
\label{eq:qq3}
\end{eqnarray}
The correlation function in the path integral formalism is computed by
inserting the components of fields on the forward and return path of
the contour.  The components of the matrix Green's functions that we
will be interested in are called greater (lesser) Green's functions,
denoted as $G^{>(<)}(t_1,t_2)$, and are defined in the following
manner \footnote{We use the commutation relation
  $\left\{\psi_i,\psi_j\right\}=\delta_{ij}$. So, $G^{>}(t,t)=-i/2$
  and $G^{<}(t,t)=i/2$.}
\begin{equation}
  \label{eq:qq4}
  \begin{split}
    G^{>}(t_1,t_2)\equiv G(t_1^{-},t_2^{+})&=-i\langle
    \psi_i(t_2)\psi_i(t_1)\rangle \\
  G^{<}(t_1,t_2)\equiv
  G(t_1^{+},t_2^{-})&=i\langle
  \psi_i(t_1)\psi_i(t_2)\rangle\ ,
  \end{split}
\end{equation}
where by $t_i^{+}$ we mean $t_i$ on the upper contour and $t_i^{-}$
denotes $t_i$ on the lower contour, and the contracted index $i$
simply denotes a sum over $i$. The relative minus sign above is due to
swapping of the position of two Majorana fermions under contour
ordering.  From the above definitions, for Majorana fermions,
\begin{equation}
G^{<}(t_2,t_1)=-G^{>}(t_1,t_2)\
\label{grleG}
\end{equation}
This relation holds even for non-equilibrium dynamics
\cite{Eberlein:2017wah, PhysRevX.5.041005}.

This model exhibits conformal symmetry in the infrared which is
spontaneously broken by the $h=2$ mode, where $h$ is the quantum
number of the $SL(2)$ subgroup of the conformal symmetry.  This $h=2$
mode has chaotic behaviour for $q\ge 4$.  It turns out that the $h=2$
mode saturates the chaos bound
$\lambda_L = 2\pi/\beta$\cite{Maldacena:2015waa}.  The model with only
$q=2$ term, however, does not have chaotic behaviour.  This is clearly
due to the quadratic nature of the action and as a result the model
is integrable.  We are interested in studying the SYK model with time
dependent coupling which can exhibit different behaviour by virtue of
having the coupling as a function of time.

Our main object of interest is the Kadanoff-Baym equations
which we will use to analyse the non-equilibrium dynamics of the SYK
model.  Before we set up the Kadanoff-Baym equations, let us consider
the Schwinger-Dyson equation.

\subsection{The Schwinger-Dyson(SD) equations}

We will consider the time dependent Hamiltonian which describes
different quench protocols depending on the kind of time dependence we
allow for the couplings of the theory.  To simplify the matter we will
extract the time dependence of the couplings and write it in terms of
separate functions of time.  For example, up to the quartic fermion
interaction {\em i.e.}, $q=4$, the Hamiltonian is
\begin{equation}\label{eq:qq5}
H(t)=i\sum_{i<j}J_{2,ij}f_2(t)\psi_i\psi_j
-\sum_{i<j<k<l}J_{4,ijkl}f_4(t)\psi_i\psi_j\psi_k\psi_l\ ,
\end{equation}
where, $f_2(t)$ and $f_4(t)$ contain the time dependence of the
couplings.  The partition function of this model is written in terms
of the action functional,
\begin{equation}
  \label{eq:qq6}
  S[\psi]=\int_{\mathcal{C}}dt\left\lbrace
    \frac{i}{2}\sum_i\psi_i\partial_t\psi_i-i\sum_{i<j}J_{2,ij}f_2(t)\psi_i\psi_j
    +\sum_{i<j<k<l}J_{4,ijkl}f_4(t)\psi_i\psi_j\psi_k\psi_l\right\rbrace\ .
\end{equation}
All the interaction terms in the SYK model couple all fermions to each
other and have random couplings.  The randomness of the coupling is
meant to mimic the disorder in the system.  We will average the
partition function over the gaussian distributed random couplings,
\begin{equation}\label{eq:qq7}
  Z=\int \mathcal{D}\psi\int\mathcal{D}J_{2,ij}\int\mathcal{D}J_{4,ijkl}\
  \mathcal{P}_1(J_{2,ij})\mathcal{P}_2(J_{4,ijkl})\exp(iS[\psi])\ ,
\end{equation}
where the gaussian weight functions, $\mathcal{P}_1(J_{2,ij})$ for the
quadratic coupling and $\mathcal{P}_2(J_{4,ijkl})$ for the
quartic coupling have width $2J_2^2/N$ and $12J_4^2/N$ respectively.
Usually in the quenched disorder the integration over the random
variables is carried out at the end of the computation, however, in
the large $N$ limit we can reverse the order.  Carrying out the
gaussian integral over the quadratic and quartic couplings gives us
the effective action
\begin{equation}
  \label{eq:qq8}
  \begin{split}
    iS_{\rm eff} &=
    -\int_{\mathcal{C}}dt\frac{1}{2}\sum_i\psi_i\partial_t\psi_i
-\frac{1}{2}\times\frac{J_2^2}{2N}\int dt_1
dt_2\sum_{i,j}f_2(t_1)f_2(t_2)
\psi_{i}(t_1)\psi_{i}(t_2)\psi_{j}(t_1)\psi_{j}(t_2)\\
&+\frac{3J_4^2}{4! N^3}\int dt_1dt_2\sum_{i,j,k,l}f_4(t_1)f_4(t_2)
\psi_{i}(t_1)\psi_{i}(t_2)\psi_{j}(t_1)\psi_{j}(t_2)
\psi_{k}(t_1)\psi_{k}(t_2)\psi_{l}(t_1)\psi_{l}(t_2)\ .
  \end{split}
\end{equation}
In this effective action the sum runs over all values of $i,j,k,l$ and
the combinatoric factors take care of the ordering of fermions in each
term.  Following \cite{Eberlein:2017wah}, we will write this effective
action in terms of auxiliary fields and convert it into a quadratic
action in terms of the fermions.  The path integral in terms of the
auxiliary functions, suggestively named as $G(t)$ and $\Sigma(t)$,
\begin{equation}
  \label{eq:qq9}
  \begin{split}
  Z&=\int\mathcal{D}\psi\,\mathcal{D}G\,\mathcal{D}\Sigma\
  \exp\left[-\int_{\mathcal{C}}dt\frac{1}{2}\sum_i\psi^i\partial_t\psi^i
    +\frac{J_2^2N}{4}\int_{\mathcal{C}}
    dt_1dt_2 f_2(t_1)f_2(t_2)G(t_1,t_2)^2\right. \\
  &\left.\hspace{3cm} -\frac{3J_4^2N}{4!}\int_{\mathcal{C}} dt_1 dt_2
    f_4(t_1)f_4(t_2)G(t_1,t_2)^4\right.\\
  &\left. \hspace{3cm} +\frac{i}{2}\int_{\mathcal{C}} dt_1dt_2
  \Sigma(t_1,t_2)\left(G(t_1,t_2)+\frac{i}{N}\sum_i\psi_i(t_1)\psi_i(t_2)\right)\right]\ ,
  \end{split}
\end{equation}
where,
\begin{equation}\label{eq:qq10}
G(t_1,t_2)=-\frac{i}{N}\sum_i\psi_i(t_1)\psi_i(t_2)\ .
\end{equation}
The auxiliary field $\Sigma$ is introduced  so that we can implement
the constraint \eqref{eq:qq10} as an equation of motion of $\Sigma$.
This is done by implementing the constraint through the $\delta$-function.
This procedure reduces the action \eqref{eq:qq9} to quadratic form in
terms of the fermions.  We can now integrate out the Majorana fermions
and write the effective action $S[G,\Sigma]$ purely in terms of $G$
and $\Sigma$,
\begin{equation}
  \label{eq:qq13}
  \begin{split}
    S[G,\Sigma]&=-\frac{iN}{2}\Tr(\log\left[-i(G_0^{-1}-\Sigma)\right])
    +\frac{iJ_2^2N}{4}\int
    dt_1\int dt_2 f_2(t_1)f_2(t_2)G(t_1,t_2)^2\\
    &-\frac{3iJ_4^2N}{4!}\int dt_1\int dt_2
    f_4(t_1)f_4(t_2)G(t_1,t_2)^4+\frac{iN}{2}\int dt_1 dt_2
    \Sigma(t_1,t_2)G(t_1,t_2)\ .
  \end{split}
\end{equation}
An advantage of this form of the effective action is that the
Schwinger-Dyson equations can be derived as equations of motion of
this action,
\begin{gather}
\label{eq:qq14}
\Sigma(t_1,t_2)=G_0^{-1}(t_1,t_2)-G^{-1}(t_1,t_2)\\
\Sigma(t_1,t_2) =
  J_2^2f_2(t_1)f_2(t_2)G(t_1,t_2)-J_4^2f_4(t_1)f_4(t_2)G(t_1,t_2)^3 \
\label{eq:qq15}
\end{gather}
A similar analysis can be carried out for the six and higher fermion
interactions in an analogous manner.  Let us now consider the
eq.\eqref{eq:qq15} and take the convolution product with $G(t_1,t_2)$
from both right and left, this procedure gives us two equation,
\begin{eqnarray}
  \int_{\mathcal{C}}dt_3G_0^{-1}(t_1,t_3)G(t_3,t_2)=\delta_{\mathcal{C}}
  (t_1,t_2)+\int_{\mathcal{C}}dt_3\Sigma(t_1,t_3)G(t_3,t_2)\ ,\label{eq:qq16}
\\
  \int_{\mathcal{C}}dt_3G(t_1,t_3)G_0^{-1}(t_3,t_2)=\delta_{\mathcal{C}}
  (t_1,t_2)
  +\int_{\mathcal{C}}dt_3G(t_3,t_2)\Sigma(t_1,t_3)\ . \label{eq:qq17}
\end{eqnarray}
To study the Kadanoff-Baym equations besides
eq. \eqref{eq:qq16}, \eqref{eq:qq17} we will need the retarded, the
advanced and the Keldysh Green's functions which are defined as
\begin{eqnarray}
G^R(t_1,t_2) &\equiv &
                       \Theta(t_1-t_2)[G^{>}(t_1,t_2)-G^{<}(t_1,t_2)]\
                       ,
                       \label{eq:qq18}\\
G^A(t_1,t_2) &\equiv &
                       \Theta(t_2-t_1)[G^{<}(t_1,t_2)-G^{>}(t_1,t_2)]\
                       ,
                       \label{eq:qq19}\\
  G^K(t_1,t_2) &\equiv & G^{>}(t_1,t_2)+G^{<}(t_1,t_2)\ .
                         \label{eq:qq20}
\end{eqnarray}
Along these lines define the retarded, advanced self-energy in the
following manner.
\begin{eqnarray}
\Sigma^R(t_1,t_2) &\equiv &
                            \Theta(t_1-t_2)[\Sigma^{>}(t_1,t_2)-\Sigma^{<}(t_1,t_2)]\
                            ,\label{eq:qq21} \\ 
\Sigma^A(t_1,t_2) &\equiv &
                    -\Theta(t_2-t_1)[\Sigma^{>}(t_1,t_2)-\Sigma^{<}(t_1,t_2)]\
                            .\label{eq:qq22} 
\end{eqnarray}
In the next subsection we will use these ingredients to derive the
Kadanoff-Baym equations.

\subsection*{The Kadanoff-Baym (KB) equations}

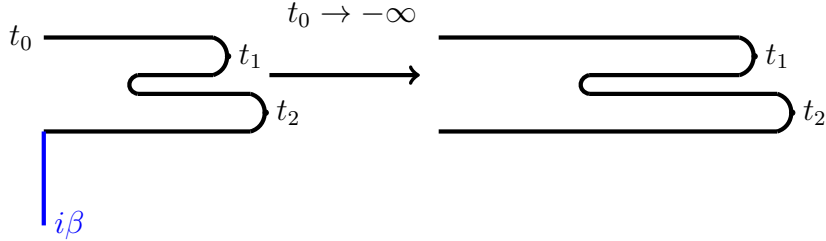
\begin{figure}[h!]
  \centering
\begin{tikzpicture}[scale=0.50]
\filldraw [black] (-3.5,1) circle (0pt) node[anchor=east]{$t_0$} ;
\draw[black, ultra thick] (-3.5,1) -- (1,1) ;
\draw[black, ultra thick] (1,1) .. controls (1.5, 0.85) and (1.5,0.15) .. (1,0) ;
\filldraw [black] (1.4,0.5) circle (2pt) node[anchor=west]{$t_1$} ;
\draw[black, ultra thick] (-1, 0) -- (1, 0) ;
\draw[black, ultra thick] (-1, 0) .. controls (-1.3, -0.08) and (-1.3,-0.44) .. (-1.,-0.5) ;
\draw[black, ultra thick] (-1, -0.5) -- (2, -0.5) ;
\draw[black, ultra thick] (2, -0.5) .. controls (2.5, -0.65) and (2.5,-1.35) .. (2, -1.5) ;
\draw[black, ultra thick] (2, -1.5) -- (-3.5, -1.5) ;
\filldraw[black] (2.4,-1) circle (2pt) node[anchor=west]{$t_2$} ;
\draw[blue, ultra thick] (-3.5,-1.5) -- (-3.5,-4) ;
\filldraw[blue] (-3.5,-4) circle (0pt) node[anchor=west]{$i\beta$} ;
\draw[ultra thick, ->] (2.5,0) -- (6.5,0) ;
\filldraw [black] (4.7,1) circle (0pt) node[anchor=south]{$t_0\rightarrow -\infty$} ;
\draw[black, ultra thick] (7,1) -- (15,1) ;
\draw[black, ultra thick] (15,1) .. controls (15.5, 0.85) and (15.5,0.15) .. (15,0) ;
\filldraw [black] (15.4,0.5) circle (2pt) node[anchor=west]{$t_1$} ;
\draw[black, ultra thick] (11, 0) -- (15, 0) ;
\draw[black, ultra thick] (11, 0) .. controls (10.7, -0.08) and (10.7,-0.44) .. (11,-0.5) ;
\draw[black, ultra thick] (11, -0.5) -- (16, -0.5) ;
\draw[black, ultra thick] (16, -0.5) .. controls (16.5, -0.65) and (16.5,-1.35) .. (16, -1.5) ;
\draw[black, ultra thick] (16, -1.5) -- (7, -1.5) ;
\filldraw[black] (16.4,-1) circle (2pt) node[anchor=west]{$t_2$} ;
\end{tikzpicture}
\caption{Contour deformation for Bogoliubov principle of weakening
  correlations.} \label{fig:condef}
\end{figure}

Equations \eqref{eq:qq16} and \eqref{eq:qq17} can be manipulated
using the real space representation of $G_0^{-1}$ on the left hand
side and contour deformation on the right hand side to write
\begin{equation}
  \label{eq:QQ-SYK:4}
  i\partial_{t_1}G^{>}(t_1,t_2)=\int_{-\infty}^{\infty}dt_3\lbrace
  \Sigma^R(t_1,t_3)G^{>}(t_3,t_2)+\Sigma^{>}(t_1,t_3)G^A(t_3,t_2)\rbrace\ .
\end{equation}  
 
\begin{equation}
  \label{eq:QQ-SYK:5}
-i\partial_{t_2}G^{>}(t_1,t_2)=\int_{-\infty}^{\infty}dt_3\lbrace
G^R(t_1,t_3)\Sigma^{>}(t_3,t_2)+G^{>}(t_1,t_3)\Sigma^A(t_3,t_2)\rbrace\ .
\end{equation}
Note that the contour starts from some time $t_0$ and the operators
are inserted in the correct order for different values of $t_1$ and
$t_2$ and then comes back to $t_0$.  For quenches starting from a
thermal state, the contour further goes down in the imaginary time
direction for an interval of length $\beta_i$ which is the inverse
temperature of the initial thermal state (Figure \ref{fig:condef}).

If one takes the limit $t_0\rightarrow -\infty$ then for all
observables at finite time, the contribution from the imaginary time
interval can be neglected which follows from the Bogoliubov principle
of weakening correlations
\cite{maciejko2007introduction}.\footnote{For this work, the
  calculation is further simplified because the free part of the
  Hamiltonian is zero.}

We will briefly explain derivation of \eqref{eq:QQ-SYK:4} using the
Langreth rules below.
Derivation of \eqref{eq:QQ-SYK:5} follows in an analogous manner.  The
left hand side of \eqref{eq:QQ-SYK:4} can be derived starting from the
equation\eqref{eq:qq16}, and choosing the Green's function
$G(t_3, t_2)$ to be the greater Green's function
$G^>(t_3, t_2)$, and integrating by parts to get
\begin{equation}
  \label{eq:QQ_SYK:1}
  \begin{split}
      \hbox{\rm L.H.S.} & = i\int_{\mathcal C} dt_3
      (\partial_{t_1}\delta_\mathcal{C}(t_1, t_3))G^>(t_3, t_2) \\
      & = i\int_{\mathcal C} dt_3
      \delta_\mathcal{C}(t_1, t_3) \partial_{t_3}G^>(t_3, t_2) \\
      & = i \partial_{t_1}G^>(t_1, t_2) \ ,
  \end{split}
\end{equation}
where we have used the fact that $G_0^{-1}$ is given by the derivative
of the $\delta$-function.  The right hand side of \eqref{eq:qq16} is
\begin{equation}
  \label{eq:QQ_SYK:2}
  \hbox{\rm R.H.S.} = \int_{\mathcal C} dt_3\Sigma(t_1^+,
  t_3)G(t_3, t_2^+)\ .
\end{equation}

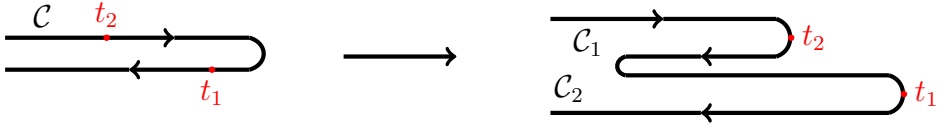
\begin{figure}[h!]
  \centering
  \begin{tikzpicture}[scale=0.50]
    \filldraw [black] (-4.5,0.5) circle (0pt) node[anchor=south]{$\mathcal{C}$} ;
\draw[ultra thick, ->] (-5.5,0.5) -- (-1.0,0.5) ;
\draw[black, ultra thick] (-1.0,0.5) -- (1,0.5) ;
\draw[black, ultra thick] (1,0.5) .. controls (1.5, 0.35) and (1.5,-0.2) .. (1,-0.35) ;
\draw[ultra thick, ->] (1, -0.35) -- (-2.2, -0.35) ;
\draw[black, ultra thick] (-2.2, -0.35) -- (-5.5, -0.35) ;
\filldraw [red] (0.0,-0.35) circle (2pt) node[anchor=north]{$t_1$} ;
\filldraw[red] (-2.8,0.5) circle (2pt) node[anchor=south]{$t_2$} ;
\draw[ultra thick, ->] (3.5,0) -- (6.5,0) ;
\draw[ultra thick, ->] (9,1) -- (12.0,1) ;
\draw[black, ultra thick] (12,1) -- (15,1) ;
\draw[black, ultra thick] (15,1) .. controls (15.5, 0.85) and (15.5,0.15) .. (15,0) ;
\filldraw [red] (15.4,0.5) circle (2pt) node[anchor=west]{$t_2$} ;
 \filldraw [black] (10,1) circle (0pt) node[anchor=north]{$\mathcal{C}_1$} ;
\draw[ultra thick, ->] (15,0) -- (13,0) ;
\draw[black, ultra thick] (13, 0) -- (11, 0) ;
\draw[black, ultra thick] (11, 0) .. controls (10.7, -0.08) and (10.7,-0.44) .. (11,-0.5) ;
\draw[black, ultra thick] (11, -0.5) -- (18, -0.5) ;
\draw[black, ultra thick] (18, -0.5) .. controls (18.5, -0.65) and
(18.5,-1.35) .. (18, -1.5) ;
\draw[ultra thick, ->] (18,-1.5) -- (13.0,-1.5) ;
\draw[black, ultra thick] (13, -1.5) -- (9, -1.5) ;
\filldraw[red] (18.4,-1) circle (2pt) node[anchor=west]{$t_1$} ;
\filldraw [black] (9.5,-1.5) circle (0pt) node[anchor=south]{$\mathcal{C}_2$} ;
\end{tikzpicture}
\caption{Contour deformation for Langreth Rules.} \label{fig:condefL}
\end{figure}
Using the contour deformation we can rewrite \eqref{eq:QQ_SYK:2} as
\begin{equation}
  \label{eq:QQ_SYK:3}
  \int_{\mathcal{C}}dt_3\Sigma(t_1^{+},t_3)G(t_3,t_2^{+})= \int_{\mathcal{C}_1}
  d\tau\Sigma(t_1,\tau)G^{>}(\tau,t_2)
  +\int_{\mathcal{C}_2}dt\Sigma^{>}(t_1,t)G(t,t_2)\ .
\end{equation}
The first term in \eqref{eq:QQ_SYK:3} can be written as
\begin{equation}
  \label{eq:QQ_SYK:4}
  \begin{split}
    \int_{\mathcal{C}_1}d\tau\Sigma(t_1,\tau)G^{>}(\tau,t_2)
    &=\int_{-\infty}^{t_1}\!\!d\tau 
\Sigma^{>}(t_1,\tau)G^{>}(\tau,t_2)+\int_{t_1}^{-\infty}\!\!d\tau
\Sigma^{<}(t_1,\tau)G^{>}(\tau,t_2)\\
&\hspace{-2.5cm}=\int_{-\infty}^{\infty}d\tau\Theta(t_1-\tau)\Sigma^{>}(t_1,\tau)
G^{>}(\tau,t_2)-\int_{0}^{\infty}d\tilde{\tau}\Sigma^{<}(t_1,\tilde{\tau})
G^{>}(\tilde{\tau},t_2)\ ,
  \end{split}
\end{equation}
where, $\tilde\tau=t_1 - \tau$.  Inserting Heaviside
$\Theta(\tilde\tau)$ function in the term involving $\Sigma^<$ we can
extend the integration limit from $(0,\infty)$ to $(-\infty, \infty)$.
After substituting $\tilde{\tau}=t_1-\tau$, the
integral remains invariant. So we get,
\begin{equation}
  \label{eq:QQ_SYK:5}
  \begin{split}
    \int_{\mathcal{C}_1}d\tau\Sigma(t_1,\tau)G^{>}(\tau,t_2) &=
\int_{-\infty}^{\infty}
d\tau\Theta(t_1-\tau)\left(\Sigma^{>}(t_1,\tau)-
\Sigma^{<}(t_1,\tau)\right)G^{>}(\tau,t_2)\ ,\\
\int_{\mathcal{C}_1}d\tau\Sigma(t_1,\tau)G^{>}(\tau,t_2)
 &= \int_{-\infty}^{\infty}d\tau\Sigma^R(t_1,\tau)G^{>}(\tau,t_2)\ .
  \end{split}
\end{equation}
Similar manipulations can be carried out for the second term in
\eqref{eq:QQ_SYK:3} to get,
\begin{equation}
  \label{eq:QQ_SYK:6}
  \int_{\mathcal{C}_2}dt\Sigma^{>}(t_1,t)G(t,t_2)=\int_{-\infty}^{\infty}dt
 \Sigma^{>}(t_1,t)G^A(t,t_2)\ .
\end{equation}

\subsection{Eigenstate Thermalization Hypothesis}
\label{sec:eigenst-therm-hypoth}

It has been shown that the $q=4$ SYK model with Majorana fermions
\cite{Hunter-Jones:2017raw, Haque:2017bts} and complex fermions
\cite{Sonner:2017hxc} with large but finite $N$ satisfy the eigenstate
thermalization hypothesis (ETH).  Although it has been claimed
\cite{Magan:2015yoa} that $q=2$ SYK model with complex fermions
satisfies ETH, it was later found that the finite $N$ scaling in $q=2$
SYK model with Majorana fermions does not scale correctly with the
system size\cite{Haque:2017bts}.  It has therefore been suggested that
$q=4$ SYK model should thermalize while the $q=2$ model should not.
Our results do not conflict with this suggestion, however, note that
ETH necessarily involves long-time averaging of the observables
\cite{Deutsch:PhysRevA.43.2046, Srednicki:1993im,
  Rigol:PhysRevLett.108.110601}.  Long time averaging is not necessary
for thermalization or equilibration in many scenario of quantum
quenches \cite{Cramer:PhysRevLett.100.030602}, even in free theories
\cite{Mandal:2015kxi}.  In fact, it is not even clear what is the
relation of ETH with such thermalization or equilibration processes
which do not involve long-time averaging after quantum quenches. Also
note that in black hole collapse geometries
\cite{Bhattacharyya:2009uu, Caceres:2014pda, Balasubramanian:2011ur},
there is no long-time averaging invloved. These geometries are the
bulk duals of thermalization in the corresponding boundary CFT.

\subsection{Kourkoulou-Maldacena states and Instantaneous
  thermalization}
\label{sec:calabr-cardy-stat}

In this section we will introduce certain pure excited states in SYK
models.  The motivation for constructing these states comes from the
boundary state ansatz of quantum quenches in 1D systems in the
thermodynamic limit \cite{Calabrese:2005in}.  The ansatz by Calabrese
and Cardy corresponds to starting from the ground state of a gapped
theory and quenching it to a gapless theory (1+1D CFT), the final
state obtained after the quench has the generic form
\begin{equation}
|CC\rangle=e^{-\kappa H_{CFT}}|B\rangle\
\label{ccstate}
\end{equation}
where $\kappa>0$ is a parameter fixed by the quench process, $H_{CFT}$
is the Hamiltonian of the final gapless theory and $|B\rangle$ is a
conformally invariant boundary state (B state) of the CFT.  We will
refer to these states as Calabrese-Cardy(CC) states.  Determination of
the particular B state that is relevant for the description of the
post quench state of the system for a specific quantum quench is a
non-trivial problem \cite{Cardy:2017ufe}.   Nevertheless, using
conformal symmetry of the final theory, it can be shown that
expectation values of one-point and two-point functions effectively
thermalize, where the expectation values in the long-time limit are
described by a thermal ensemble with inverse temperature
$\beta=4\kappa$.  In fact, it has been shown that finite subsystems
thermalize where again the long-time limit is described by a thermal
ensemble with inverse temperature $\beta=4\kappa$ \cite{Cardy:2014rqa,
  Mandal:2015jla}.  Since the quench process started from the ground
state, the system always remains in a pure state.   An interesting
aspect of this process of thermalization of subsystems is that
correlation functions of holomorphic operators of the final CFT thermalize
instantaneously \cite{Mandal:2015kxi, Paranjape:2016iqs}.
  
We will now consider certain pure excited states in SYK models.
These states were first constructed by Kourkoulou and Maldacena in
\cite{Kourkoulou:2017zaj}.  Considering $N$ majorana fermions, the
analogous B states are defined as 
\begin{equation}
(\psi^{2k-1}-is_k\psi^{2k})|B_s\rangle=0, \qquad s_k=\pm 1, \qquad
k=1,...., N/2
\label{bs_state}
\end{equation}
Hence, there are $2^{N/2}$ number of such B states.  These are high
energy states.  One can produce lower energy states by evolving these B
states for a finite euclidean time $\kappa$.  We will refer to these
low energy states as KM states. 
\begin{equation}
|KM\rangle=e^{-\kappa H}|B_s\rangle\
\label{kmstate}
\end{equation}
An interesting feature of KM states is that, in the large $N$ limit,
``diagonal'' two-point functions $\psi^i(t_1)\psi^i(t_2)$ are
``instantaneously thermalized''(using the 1+1D CFT terminology used
above)
\begin{equation}
\langle KM|\psi^i(t_1)\psi^i(t_2)|KM\rangle=\text{Tr}\left[e^{-\beta
    H} \psi^i(t_1)\psi^i(t_2)\right], \qquad i=1,..., N \to \infty 
\end{equation}
where the effective inverse temperature $\beta=2\kappa$.  The
``off-diagonal'' two-point functions $\psi^{2k-1}(t_1)\psi^{2k}(t_2)$
have non-trivial time dependence and decay to zero in the long-time
limit.  These ``off-diagonal'' two-point functions are zero in a thermal
ensemble. The KM states also have interesting bulk duals in $AdS_2$.

Unlike in 2D CFT quenches, we could not find any quench scenario with
disordered couplings where the final state is the KM state.  This work
was initially inspired by our curiosity about the possibility of the
KM states being the final states of step quenches but not for bump
quenches in SYK models.  The negative result that the final states in
quenches in SYK models are not KM states leads to deeper understanding
of the thermalization process in chaotic theories.  We will comment
further on this issue in the concluding section \ref{sec:condis}.

\section{Quantum Quenches in SYK models}
\label{sec:quantum-quench-model}

\begin{figure}[ht!]
\begin{center}
\includegraphics[width=0.4\textwidth]{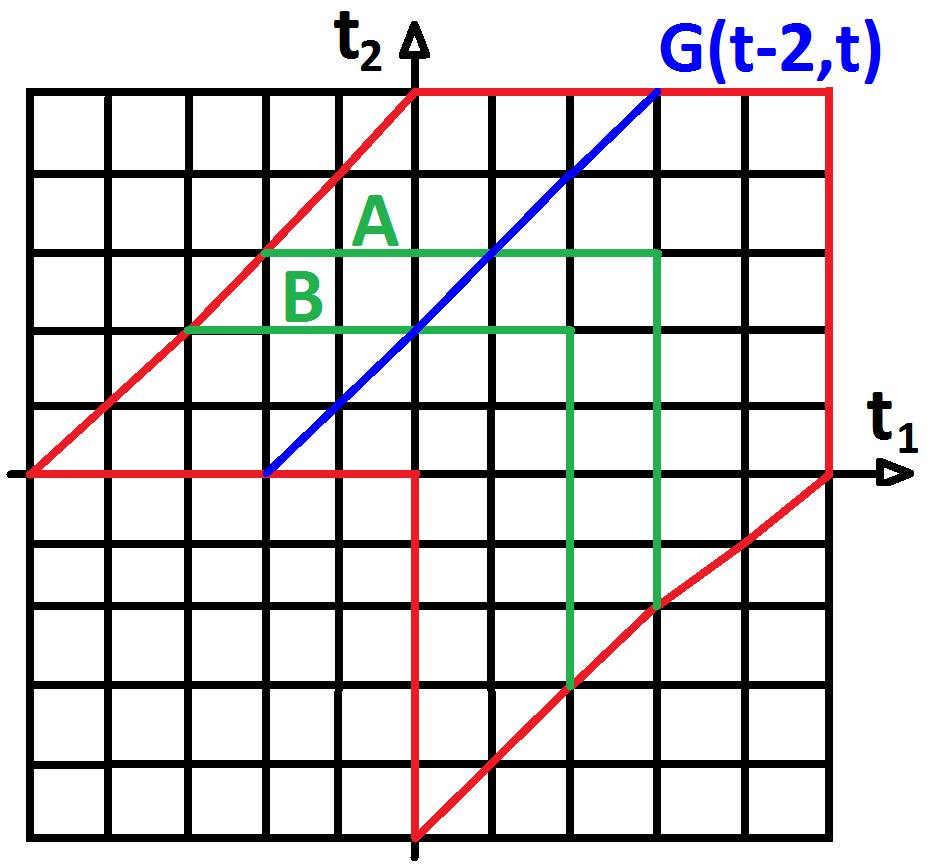}
\caption{\small The red lines mark the grid used for solving the
  Kadanoff-Baym equations. This corresponds to ignoring terms on the
  top left of the second quadrant and the bottom right of the fourth
  quadrant where the values of $G^>(t_1,t_2)$ are negligible.}
\label{fig:grid_KB}
\end{center}
\end{figure}

The KB equations are solved numerically after discretizing the two
time arguments $t_1$ and $t_2$. For quenches in $q=2$ theory, we could
start from the ground state, since the Green's function oscillates and
decays fast with time.  For all other cases, we start with a thermal
state which gives an exponential decay of the initial data as a
function of the relative time difference. Moreover, since we start
from a stationary state, all the initial data in the third quadrant
are shifted functions of the data on $(t_1<0, \, t_2=0)$ line and
$(t_1=0, \, t_2<0)$ line.  We use a grid of the kind bounded by red
coloured lines in figure \ref{fig:grid_KB}.  Since the terms far away
from the diagonal fall of exponentially fast, the grid points in the
second and fourth quadrant lying outside the red coloured lines are
ignored in our numerical code.

We used grids of three different sizes $2001\times 1001$,
$3001\times 1501$ and $4001\times 2001$ points. The computation time
grows very fast with increasing grid size. We also used a fixed time
step size $dt=0.05$.\footnote{We also checked our results with
  $dt=0.025$ to make sure some of our results are not due to finite
  size numerical time steps. But we will not present any numerical
  results of the runs with $dt=0.025$. So, $dt=0.05$ for the rest of
  the paper.} In the rest of the paper, we will suppress factors of
this time step size $dt$. So, unless it is explicitly mentioned all
the times are measured in units of $dt$.  In step protocols, the
quenches happen at $t_1=0$ and $t_2=0$.  For all the cases with bump
protocol, the perturbations\footnote{Note that we are not doing any
  perturbative or series expansion in our calculation.  The word
  `perturbation' in this context means exciting the system by turning
  on the source term which injects energy in the system.} are turned
on between $t_1=1$ and $t_1=10$, similarly between $t_2=1$ and
$t_2=10$ for the other direction.  The KB equations are solved
self-consistently in this grid using the Predictor-Corrector method.
The predicted values on line A are calculated \emph{causally} from the
data on line B as shown in figure \ref{fig:grid_KB}. The predicted
values are then corrected until the desired accuracy is obtained.

For most of quenches we are considering here, the initial data is
obtained by solving the SD equation numerically for finite inverse
temperature $\beta$ \cite{Eberlein:2017wah}.  For step quenches in
$q=2$ theory in which $J_2$ interaction is dominant, we can start from
the ground state.  The initial data are obtained by solving the SD
equation in the ground state ($\beta\to\infty$) numerically.  In this
case we use
\begin{equation}
\lim_{\beta\to\infty}\frac{1}{1+e^{-\beta \omega}} =\Theta(\omega)=\begin{cases}
    \quad 0, & \text{if} \quad \omega < 0.\\
    \quad 1/2, & \text{if} \quad\omega = 0.\\
    \quad 1, & \text{if} \quad\omega > 0.
    \end{cases}
    \label{stepfunc}
\end{equation}
In case of the bump quench in $q=2$ theory, for cases in which we
start from the ground state, the initial data is calculated using the
analytic expression for $G^>(t_1,t_2)$.  The greater Green's function
in ground state for $q=2$ theory is
\begin{equation}
G^>(t_1,t_2)=\frac{1}{2J_2 (t_1-t_2)}\left[ J_1(2J_2
  (t_1-t_2))-iH_1(2J_2 (t_1-t_2))\right]\ .
\end{equation}

\paragraph*{Calculation of final temperature:} The temperature in the
long time limit is calculated using the relation
\cite{Eberlein:2017wah}
\begin{equation}
\frac{i G^K(\omega)}{A(\omega)}=\tanh\left(\frac{\beta\omega}{2}\right)\
\label{effbeta}
\end{equation}
where $G^K(\omega)$ is the Fourier transform of the Keldysh Green's
function $G^K(t_1,t_2)$ (\ref{eq:qq20}) which is a function of only
$t_1-t_2$ in a thermal ensemble and $A(\omega)$ is
\begin{equation}
A(\omega)=-2\, \text{Im}\, G^R(\omega)\
\end{equation}
$G^R(\omega)$ is the Fourier transform of the retarded Green's
function $G^R(t_1,t_2)$ (\ref{eq:qq18}) which also is a function of
only $t_1-t_2$ in a thermal ensemble.

The relation (\ref{effbeta}) is a result of the KMS condition which
ensures \cite{kamenev2011field} that
\begin{equation}
G^>(\omega)=-e^{\beta\omega}G^<(\omega)\ ,
\end{equation}
and it holds for all fermionic theories.  We can therefore conclude
that the system under consideration has thermalized only if the
quantity on the LHS of (\ref{effbeta}) has $\tanh$ profile as a
function of the frequency $\omega$.
Note that for the determination of the final temperature we also have to
use the relation between greater and lesser Green's functions
(\ref{grleG}).

\paragraph*{Check for energy conservation:} We also check for energy
conservation to ensure that our numerical results are correct. From
(\ref{eq:qq9}), the total energy as a function of time $t_1$ is given
by
\begin{eqnarray}
E(t_1)&=&\int_\mathcal{C} dt_2 \Sigma(t_1,t_2)G(t_1,t_2)\nonumber\\
&=& \int_{-\infty}^{t_1} dt_2
    \left(\Sigma^>(t_1,t_2)G^>(t_1,t_2)-\Sigma^>(t_2,t_1)G^>(t_2,t_1)\right)\
\label{energy}
\end{eqnarray}
In the second line, the first term arises from the upper half of the
contour and the second term arises from the lower half of the
contour. We have also used (\ref{grleG}) for the second
term.
\vspace{0.5cm}

The quench processes we are considering, merely satisfying
(\ref{effbeta}) in the long time limit is not sufficient to guarantee
thermalization. This is because, as we mentioned above, all fermionic
theories at finite temperature satisfy the relation (\ref{effbeta}). So, to
check thermalization, we first calculate the final temperature using the
above relation. The SD equation of the final theory is then solved at
the calculated final temperature and in the end we check if the generated real
time two-point functions agree with the two-point functions obtained from
the quench process. 

\subsection{Quenches in $q=2$ SYK model}
\label{sec:quench-syk-model}

In this subsection we will study quantum quenches in which the final
theory is the $q=2$ SYK model, that is the model which only has 1-body
(quadratic, $J_2$) interaction. These quenches are special cases
because the two-point functions equilibrate instanteneously. From
(\ref{eq:QQ-SYK:4}, \ref{eq:QQ-SYK:5}), for $q=2$ final theory, 
\begin{equation}
\partial_{t_1}G^{>}(t_1,t_2)=-\partial_{t_2}G^{>}(t_1,t_2) \Rightarrow
G^{>}(t_1,t_2)=G^{>}(t_1+dt,t_2+dt)\ 
\label{q2_notherm}
\end{equation}
This is observed in our numerical solutions of the KB equations
below. However, note that the instanteneously equilibrated
configuration is not a thermal ensemble, so the final state cannot be
a KM state. 

Since, the initial theory is $J_2$ dominant(for step quench) or a
$q=2$ theory, we can start the quench from the corresponding ground
state. We will present here only cases in which $J_4$ interaction is
used to perform both step and bump quenches. We also found similar
results for quenches using $J_6$ and $J_8$ interactions, as we expect
from (\ref{q2_notherm}).  The results are qualitatively similar for
quenches starting from thermal state.

The value of the $J_2$ coupling is always fixed at $1$. We will
present results for step quench with initial $J_4=2$ which is suddenly
turned off at time $t=0$. For bump quench, we turn on $J_4=5$ for a
time duration of $9 \times dt = 9 \times 0.05=0.45$ from time step $t=1$
to $t=10$. This same quench parameters are used for all quenches
starting from different initial temperatures including the ones
starting from ground state.

\begin{figure}[h]
\centering
\begin{subfigure}{.5\textwidth}
  \centering
  \includegraphics[width=.9\linewidth]{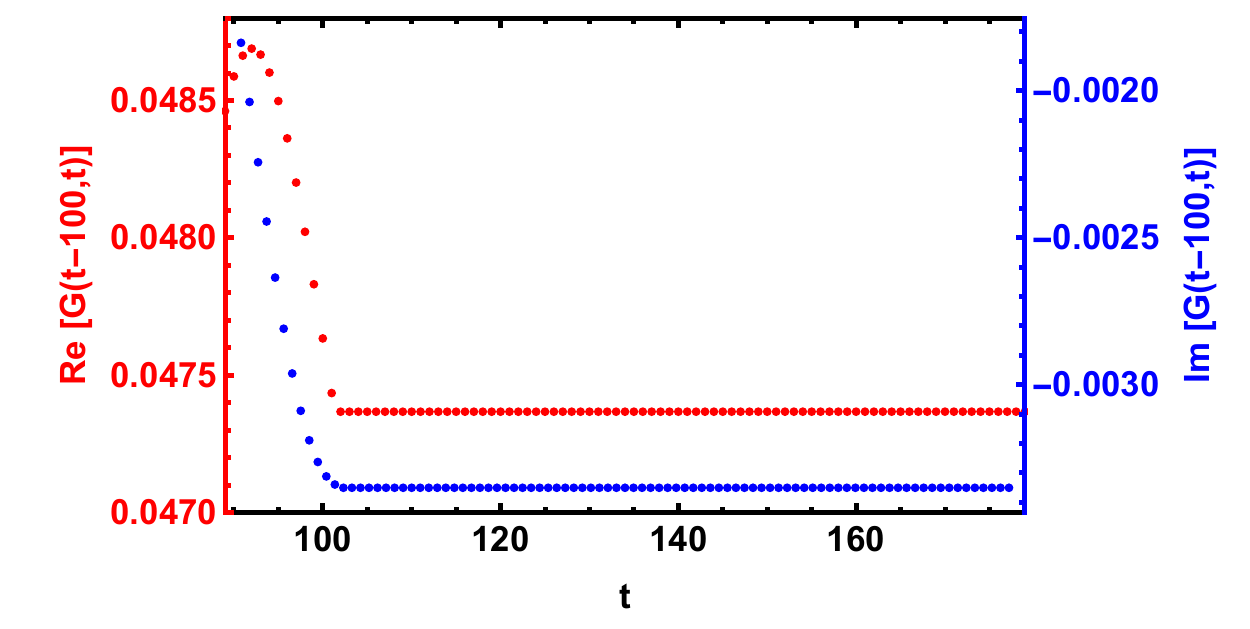}
  \caption{Step quench at $t=0$}
  \label{fig:q2qq4s}
\end{subfigure}%
\begin{subfigure}{.5\textwidth}
  \centering
  \includegraphics[width=.9\linewidth]{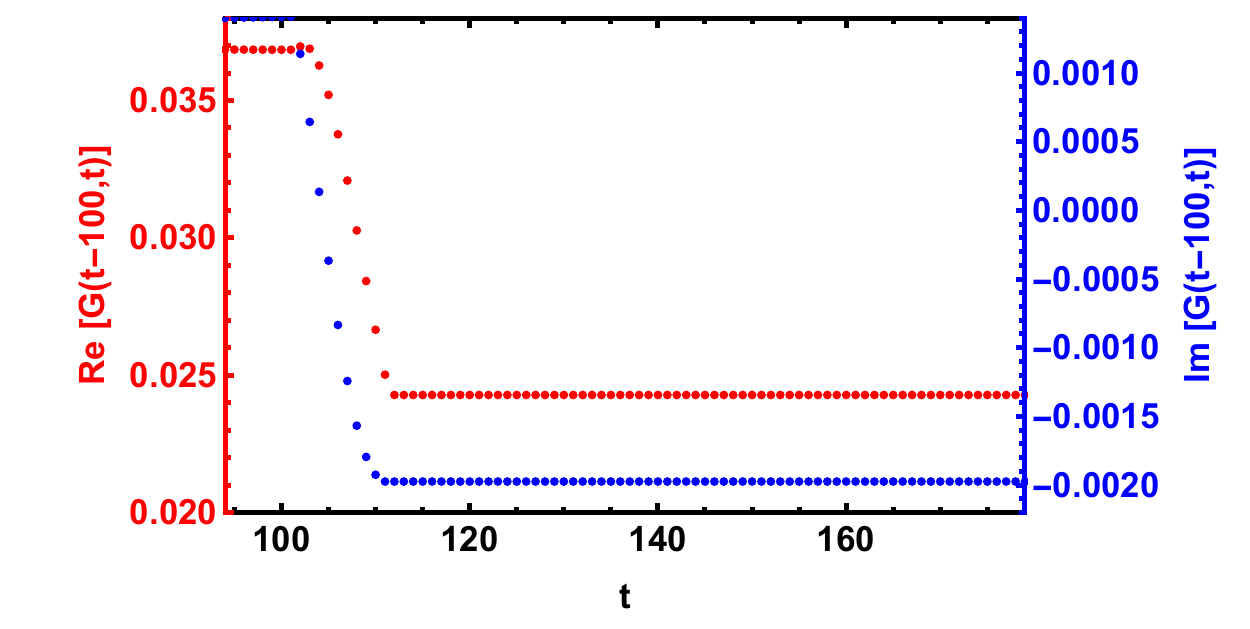}
  \caption{Bump quench between $t=1$ and $t=10$}
  \label{fig:q2qq4s}
\end{subfigure}
\caption{\small{Plots of real and imaginary parts of $G^>(t-100,t)$
    for (a) step quench, both the time arguments are outside the
    quench region for $t>100$, and for (b) bump quench, both the time
    arguments are outside the quench region for $t\geq111$. As we can
    see, the greater Green's function equilibrates
    instantaneously.}}
\label{fig:q2qq4}
\end{figure}
\begin{figure}[H]
\centering
\begin{subfigure}{.5\textwidth}
  \centering
  \includegraphics[width=.8\linewidth]{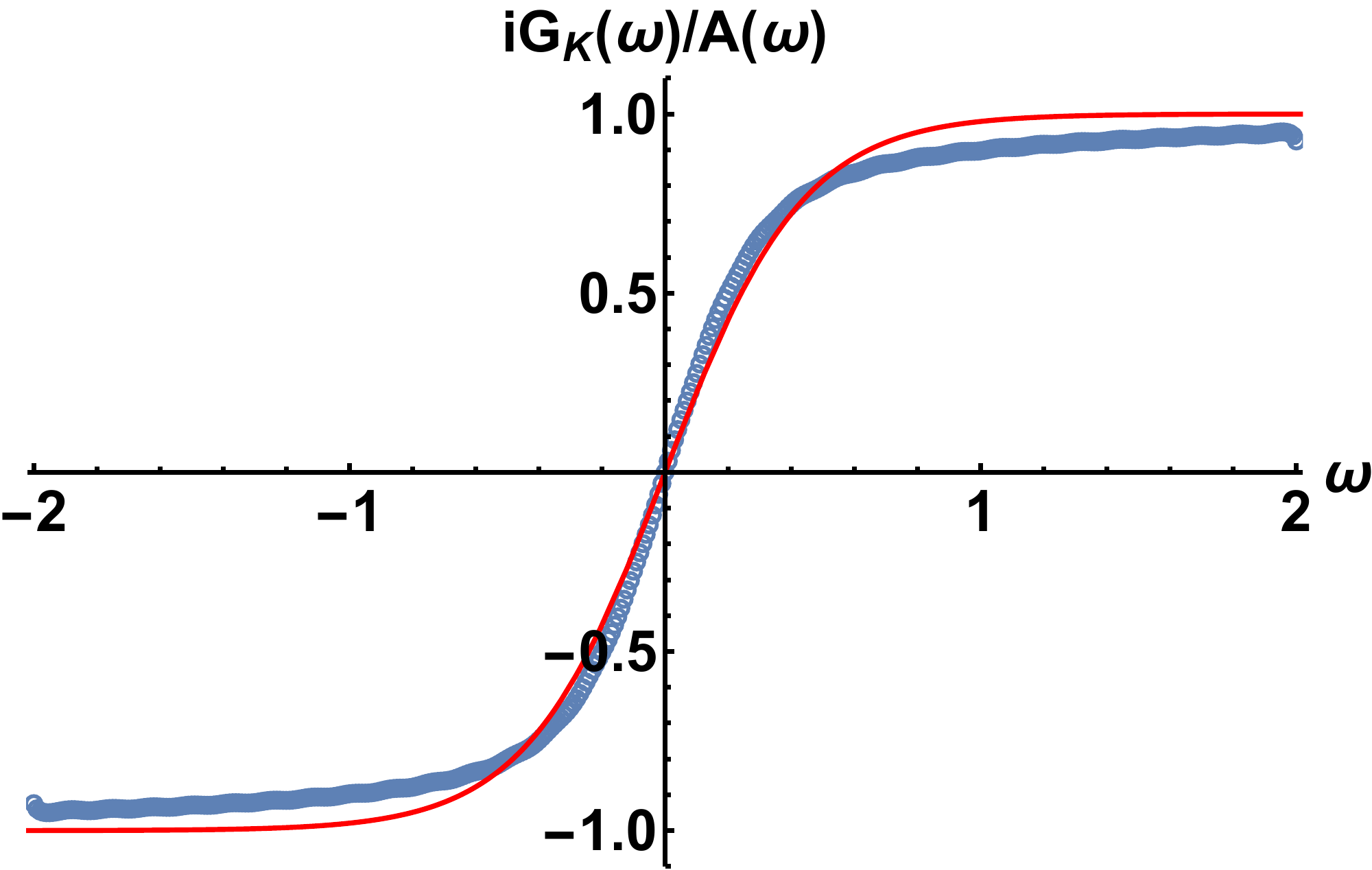}
  \caption{Step quench, $\beta_i=10$}
  \label{fig:q2qq4sb10TEMP}
\end{subfigure}%
\begin{subfigure}{.5\textwidth}
  \centering
  \includegraphics[width=.8\linewidth]{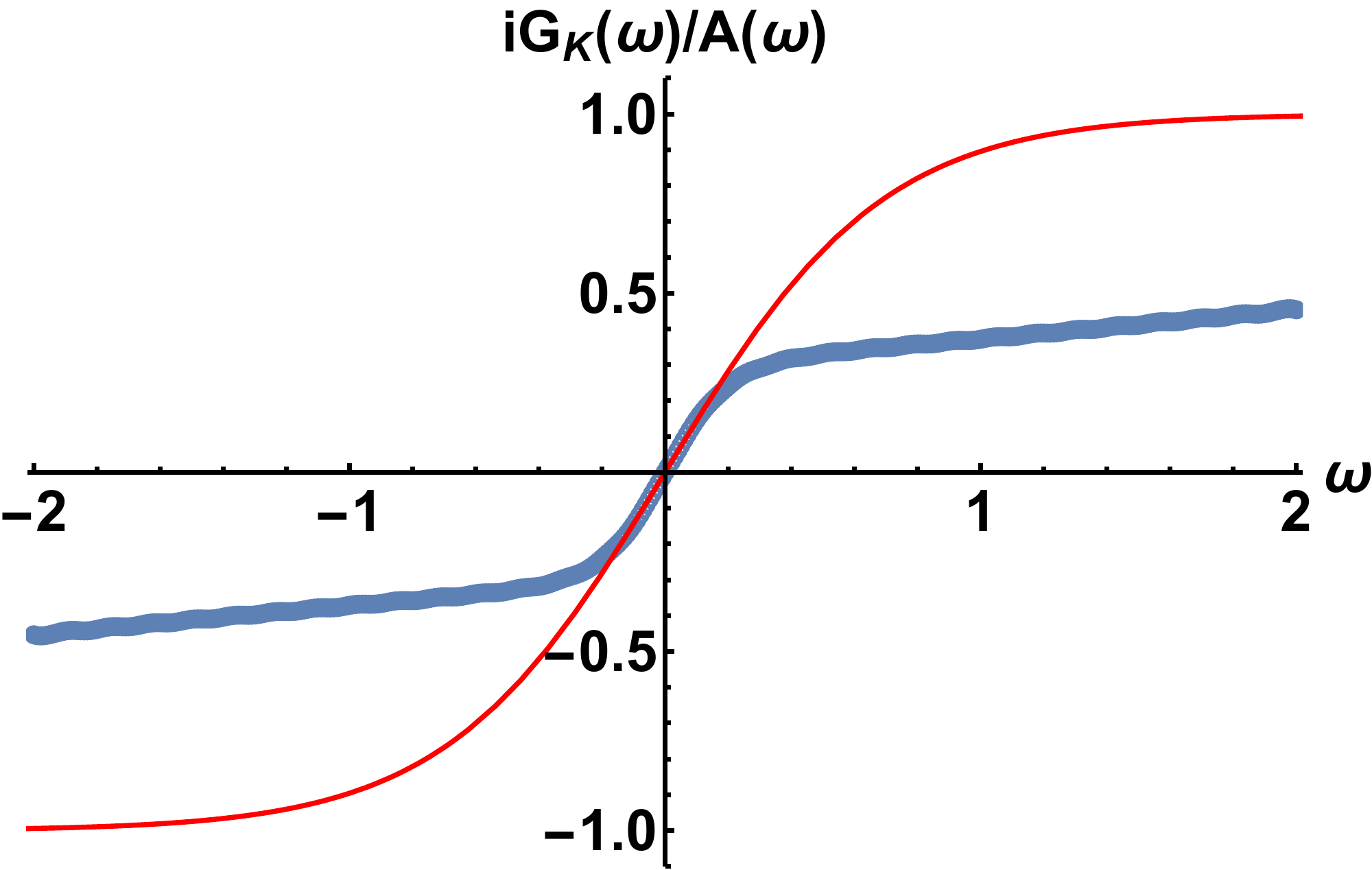}
  \caption{Bump quench, $\beta_i=10$}
  \label{fig:q2qq4bb10TEMP}
\end{subfigure}
\caption{\small{Plots of $ i G^K(\omega)/A(\omega)$ for the
    equilibrated limits of (a) step quench and (b) bump quench. For
    both the quenches, we start from a thermal state of inverse
    temperature $\beta_i=10$. The red lines are plots for the function
    $\tanh(\beta\omega/2)$ with the respective approximate $\beta_f$'s.}}
\label{fig:q2qq4b10TEMP}
\end{figure}

The step quench happens at $t=0$, the two time arguments of
$G^>(t-100,t)$ are outside the quench region if $t>100$. The bump
quench happens between $t=0$ and $t=11$ so the two time arguments are
outside the quench region if $t\geq111$.  Figure (\ref{fig:q2qq4}) are
plots of the real and imaginary parts of $G^>(t-100,t)$ as a function
of time $t$ for step and bump quenches starting from ground
states. One can see that the Green's function freezes or equilibrates
instantaneously once the two time arguments are outside the quench
regions. But the equilibrated value is different from the thermal
expectation value. Figure (\ref{fig:q2qq4b10TEMP}) compares
$ i G^K(\omega)/A(\omega)$ with $\tanh(\beta_f\omega/2)$ for step and
bump quenches starting from initial inverse temperature $\beta=10$.


\subsection{Quenches in $q=4$ SYK model}
\label{sec:quench-syk-model-2}

In this subsection we will consider quantum quenches in which the
final theory is $q=4$ SYK model which only has 2-body (quartic, $J_4$)
interaction.  We will present results for which the interaction terms
used for the quench process is $J_2$.  We also found similar results
for quenches with $J_6$ and $J_8$ interactions.  For the initial
thermal states, we considered three different inverse temperatures
$\beta_i = 10, \, 20$, and $30$.  We find that increasing the inverse
temperature from $20$ to $30$ does not affect the results much.  This
is expected since for a fairly large $\beta$, the fermion distribution
function is well represented by the step function (\ref{stepfunc}).
So, we expect that the quench starting from $\beta=20$ and $30$ should
also be qualitatively similar and quantitatively close to the quenches
starting from ground states.

Three different values of $J_4$ are used, namely, $0.5, 1$ and
$1.5$.  For step quenches, we start from a theory with $J_4$ and
$J_2$.  At $t=0$, the $J_2$ coupling is suddenly changed to $0$.  For
the bump quenches, starting from a theory with only $J_4$, $J_2$ is
turned on for a time duration of $9 \times dt = 9 \times 0.05=0.45$
from time step $t=1$ to $t=10$.  As mentioned above, we will use this
time interval for all bump quench protocol.  Changing this time
interval does not affect our main results. Longer time interval only
injects more energy into the system resulting in higher final
temperature.

\begin{figure}[h]
\centering
\begin{subfigure}{.5\textwidth}
  \centering
  \includegraphics[width=.9\linewidth]{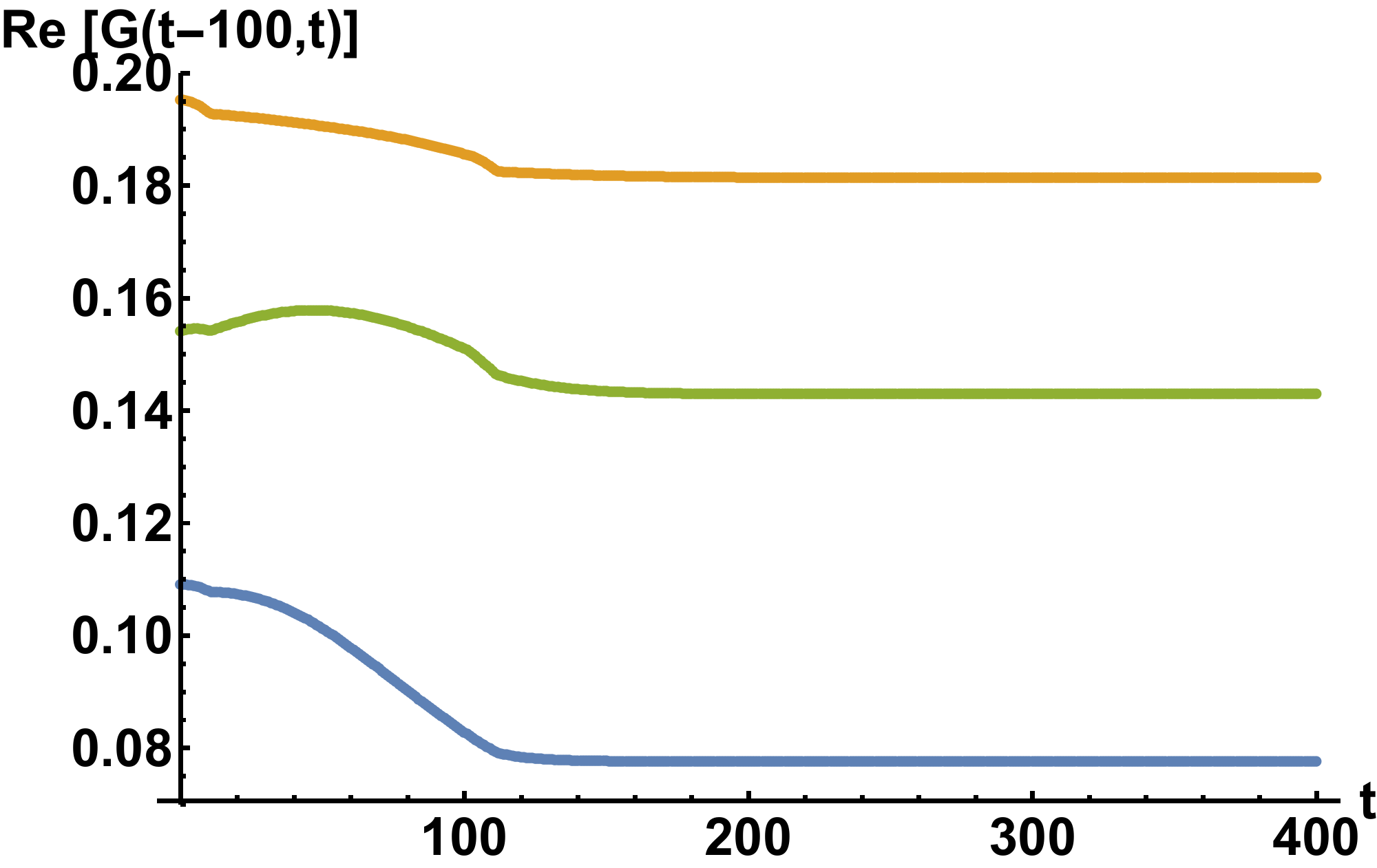}
  \caption{ }
  \label{fig:req4qq2}
\end{subfigure}%
\begin{subfigure}{.5\textwidth}
  \centering
  \includegraphics[width=.9\linewidth]{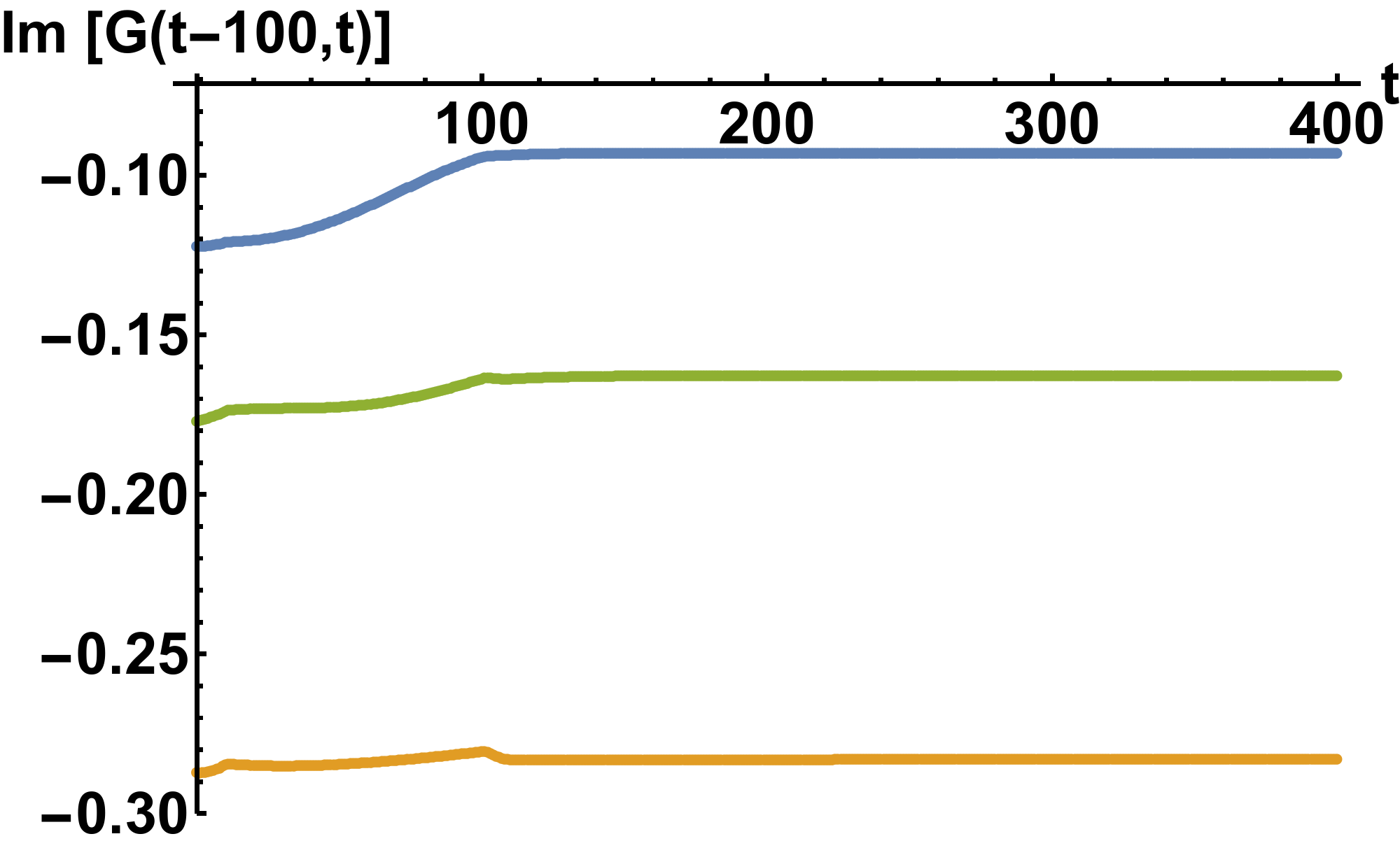}
  \caption{ }
  \label{fig:imq4qq2}
\end{subfigure}
\caption{\small{(a) Real part of the greater Green's function
    $G^>(t-100,t)$ in the SYK model with quartic interaction and
    changing the quadratic interaction $J_2$ following bump protocol
    for three different set-up using different initial temperatures
    and different values of $J_4$ and $J_2$. (b) Imaginary part of the
    same greater Green's function $G^>(t-100,t)$.}}
\label{fig:fullq4qq2}
\end{figure}
\begin{figure*}[h]
        \centering
        \begin{subfigure}[b]{0.475\textwidth}
            \centering
            \includegraphics[width=\textwidth]{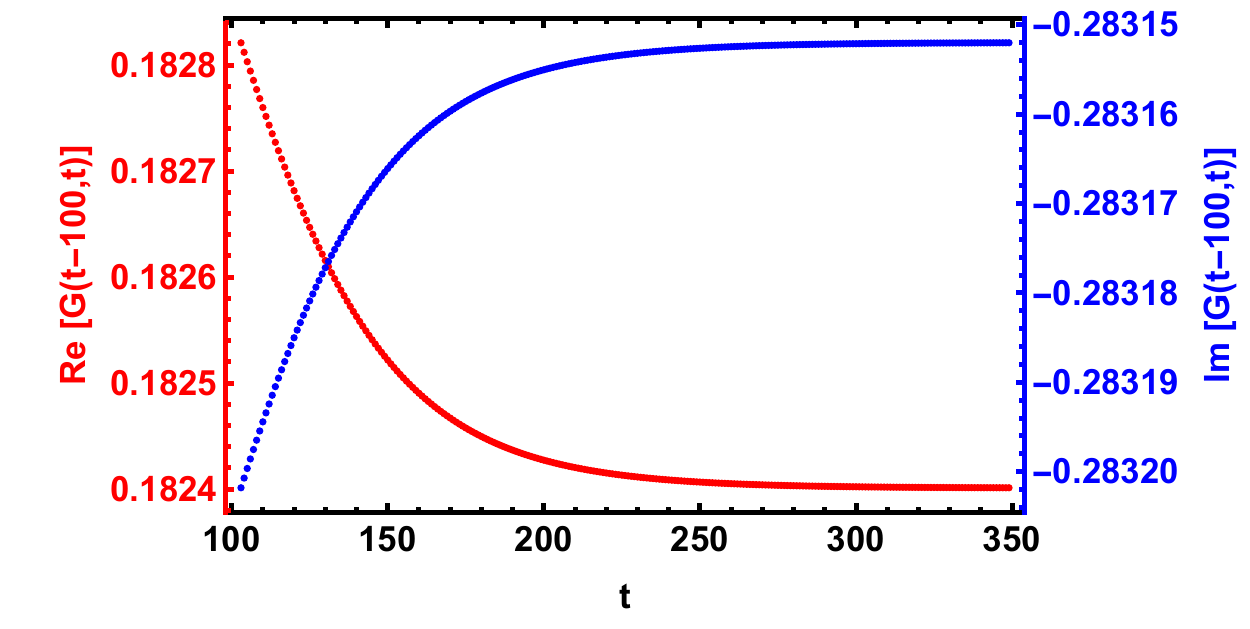}
            \caption[]%
            {{\small Step, $\beta_i=20$}}    
            \label{fig:q4qq2sb10}
        \end{subfigure}
        \hfill
        \begin{subfigure}[b]{0.475\textwidth}  
            \centering 
            \includegraphics[width=\textwidth]{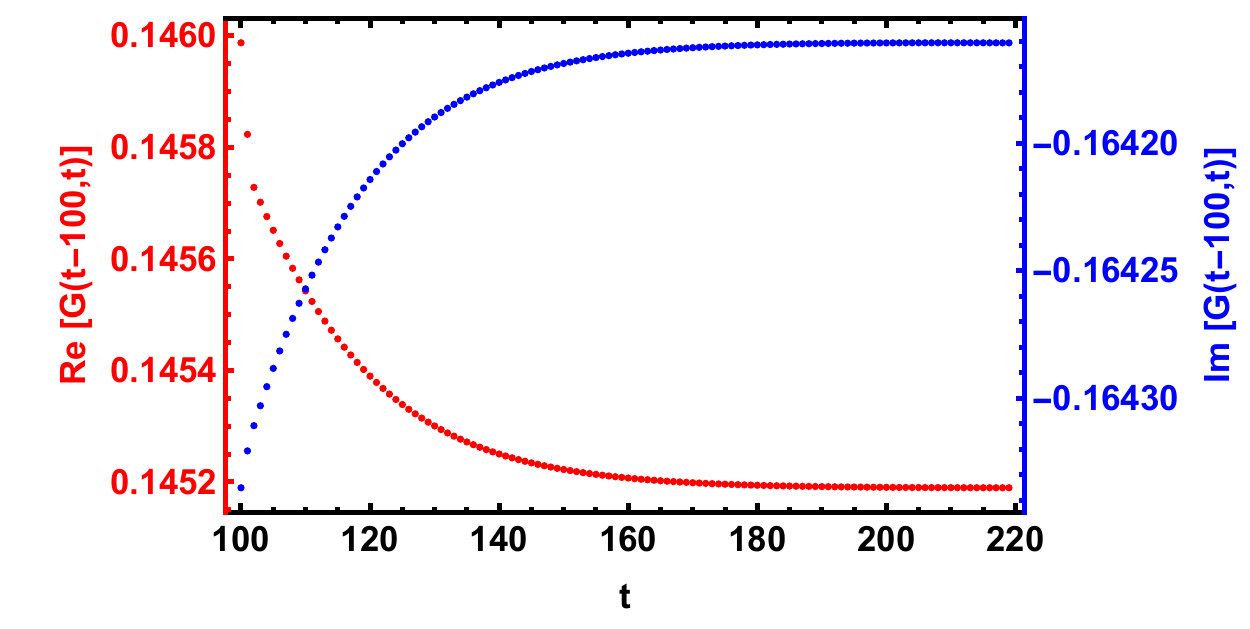}
            \caption[]%
            {{\small Step, $\beta_i=30$}}    
            \label{fig:q4qq2sb20}
        \end{subfigure}
        \vskip\baselineskip
        \begin{subfigure}[b]{0.475\textwidth}   
            \centering 
            \includegraphics[width=\textwidth]{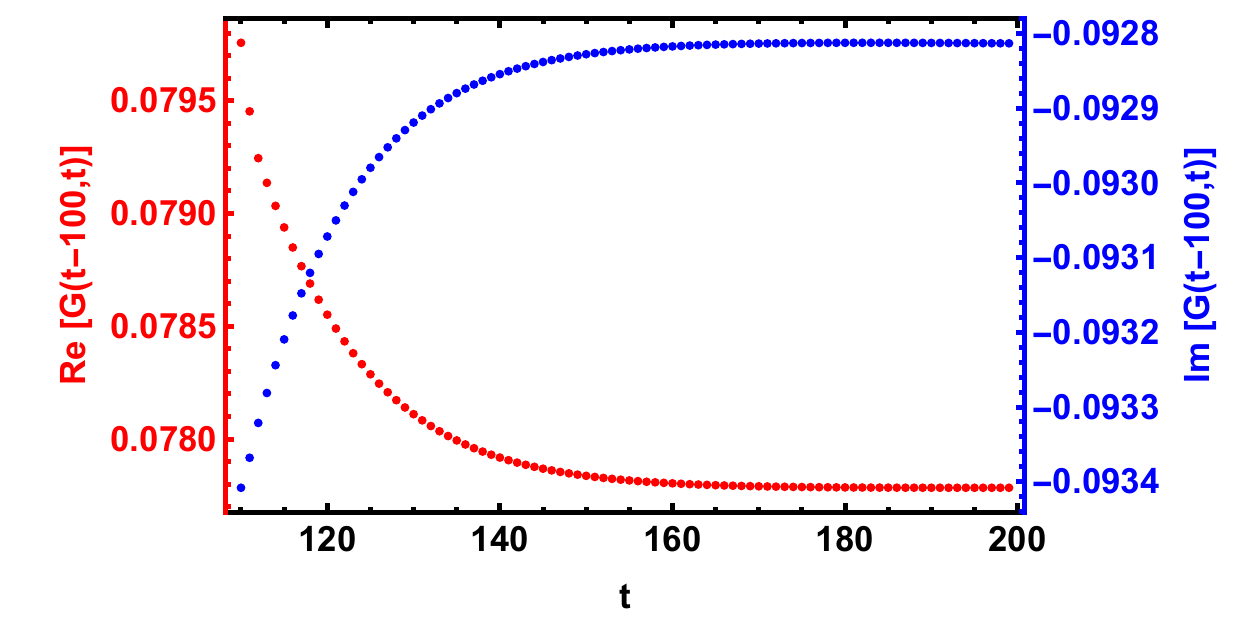}
            \caption[]%
            {{\small Bump, $\beta_i=10$}}    
            \label{fig:q4qq2bb10}
        \end{subfigure}
        \quad
        \begin{subfigure}[b]{0.475\textwidth}   
            \centering 
            \includegraphics[width=\textwidth]{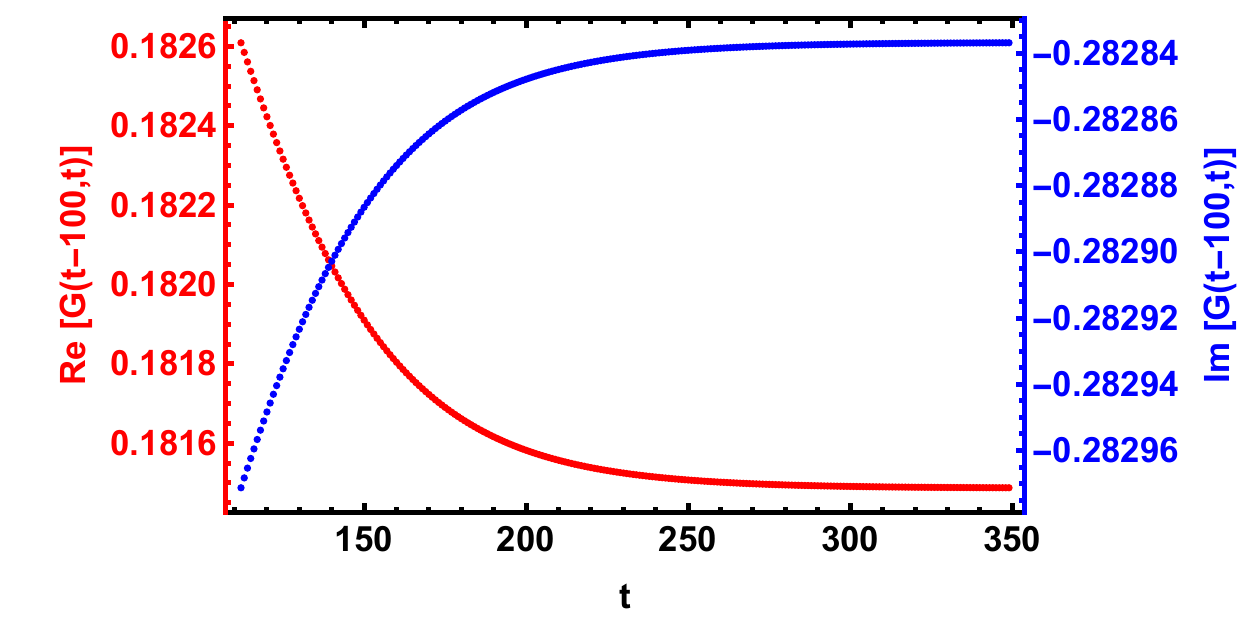}
            \caption[]%
            {{\small Bump, $\beta_i=20$}}    
            \label{fig:q4qq2bb20}
        \end{subfigure}
        \caption[]
        {Real and imaginary parts of $G^>(t-100,t)$ for different quench protocols.} 
        \label{fig:q4qq2}
\end{figure*}

Once both the time arguments are outside the quench region, we find
that the greater Green's function thermalizes rapidly but not
instantaneously, as can be seen in Figure(\ref{fig:fullq4qq2}).  Figure
(\ref{fig:q4qq2sb10}, \ref{fig:q4qq2sb20}) are two resolved plots of
$G^>(t-100,t)$ for different initial inverse temperatures as a
function of $t$ for step quenches.  Since the step quench happens at
$t=0$, both the time arguments are outside the quench region if
$t>100$.  Immediately after time $t$ crosses 100, $G^>(t-100,t)$
changes rapidly and exponentially towards its equilibrium thermal
value. The evolutions for $t>100$, both real and imaginary parts, fit
exponential functions very well.  The two exponents of the two
exponential fits for real and imaginary parts are roughly equal. This
behaviour is not a numerical artifact. The exponents do not change
with change in time step size. We have checked for different time step
sizes $dt=0.05$ and $dt=0.025$. Moreover, we have also checked energy
conservation using (\ref{energy}).

\begin{figure}[ht]
\centering
\includegraphics[width=.5\linewidth]{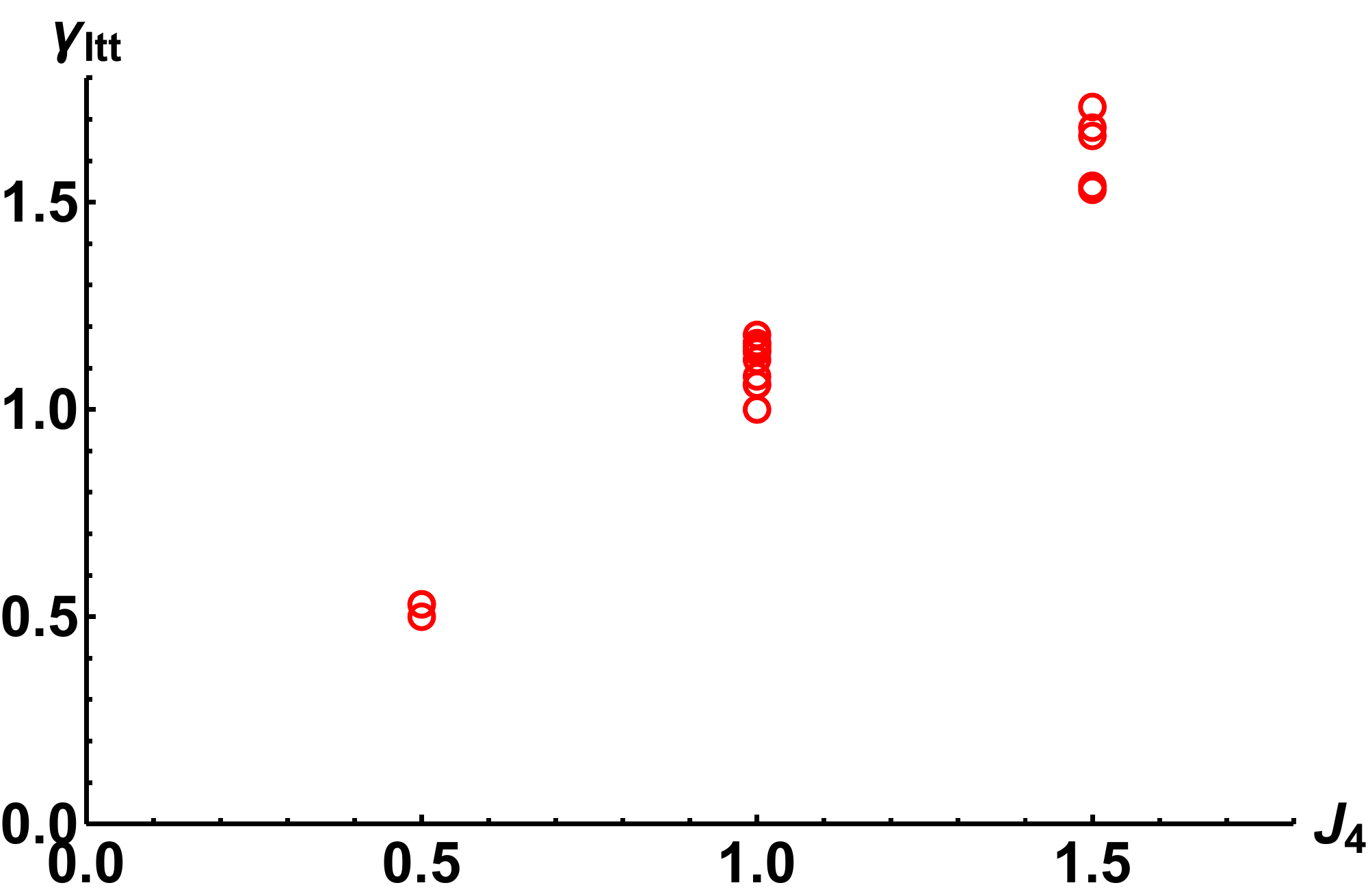}
\caption{The exponent $\gamma_{Itt}$ as a function of $J_4$.}
\label{JgammaItt}
\end{figure}

Similarly, for bump quenches in Figure (\ref{fig:q4qq2bb10},
\ref{fig:q4qq2bb20}), once the two time arguments are outside the
quench region, the Green's function thermalizes rapidly and its real
and imaginary parts fit exponential functions very well. Below, we
will consider only the exponent for the imaginary part which we will
denote by $\gamma_{Itt}$.
\begin{equation}
Im[G^>(t-100,t)]\xrightarrow{\rm{post~quench~region}} a_1+b_1 e^{-\gamma_{Itt}t}\
\end{equation}
The bump quench happens between time steps $t=0$ and $t=11$, so the
two time arguments of $G^>(t-100,t)$ are outside the quench region if
$t\geq 111$. One of the most interesting numerical result of this work
is that we find that
\begin{equation}
\gamma_{Itt}=J_4\
\label{gammaJ4}
\end{equation}
This can be seen from  Fig. (\ref{JgammaItt}) and Table \ref{table:q4}.

\begin{figure}[h]
\centering
\begin{subfigure}{.5\textwidth}
  \centering
  \includegraphics[width=.9\linewidth]{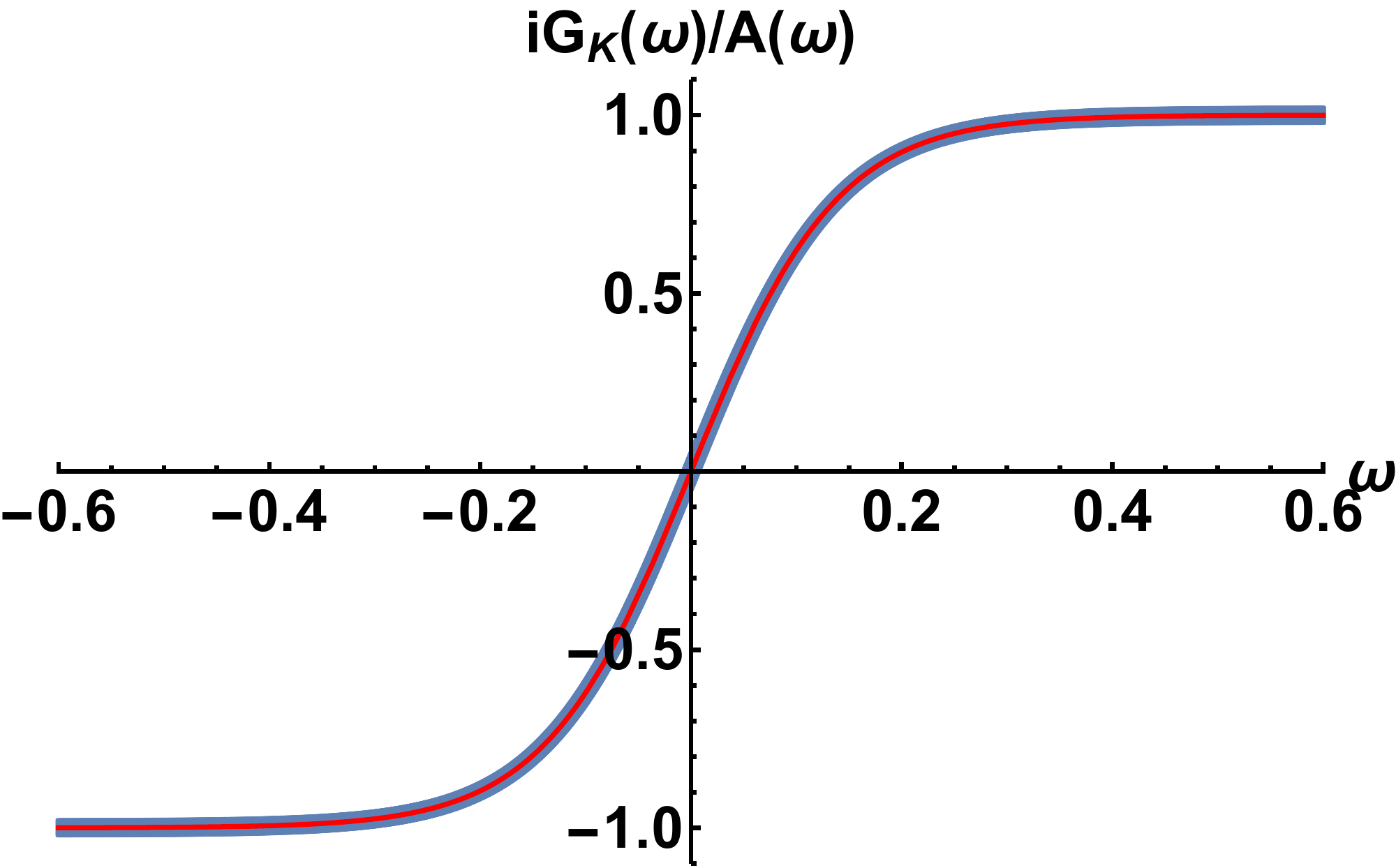}
  \caption{Step quench, $\beta_i=20$, $\beta_f=14.53$}
  \label{fig:q4qq2sb20TEMP}
\end{subfigure}%
\begin{subfigure}{.5\textwidth}
  \centering
  \includegraphics[width=.9\linewidth]{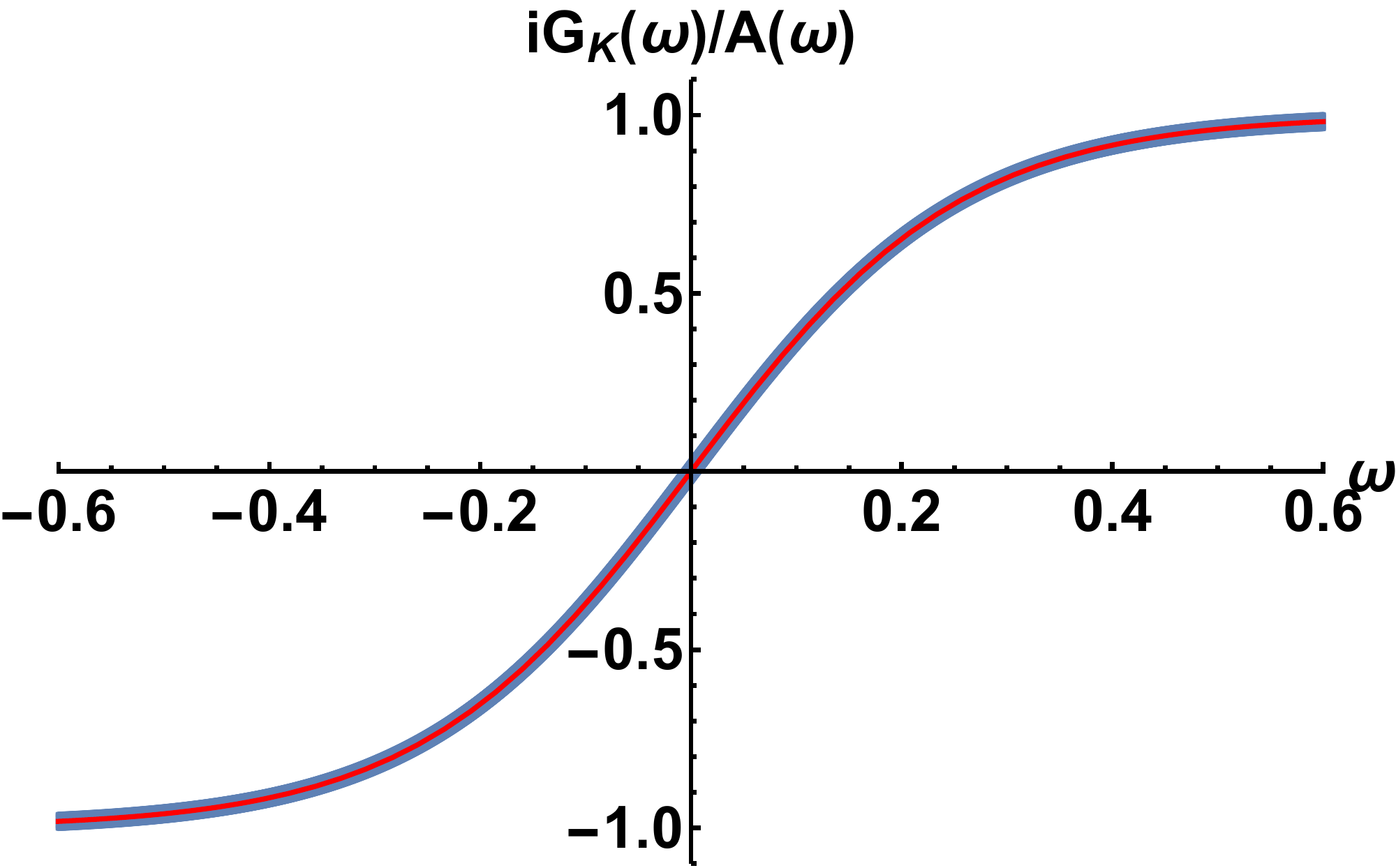}
  \caption{Bump quench, $\beta_i=20$, $\beta_f = 7.81$}
  \label{fig:q4qq2bb20TEMP}
\end{subfigure}
\caption{\small{Comparison of $ i G^K(\omega)/A(\omega)$ (blue dots)
    with $\tanh(\beta_f\omega/2)$ (thin red line).}} 
\label{fig:q4qq2b20TEMP}
\end{figure}

We also check if the final stationary limit is described by a thermal
ensemble.  For which we compare $ i G^K(\omega)/A(\omega)$ with
$\tanh(\beta_f\omega/2)$ for some final temperature $\beta_f$.  Figure
(\ref{fig:q4qq2sb20TEMP}, \ref{fig:q4qq2bb20TEMP}) are two such
comparisons.  Figure (\ref{fig:q4qq2sb20TEMP}) is for step quench with
$J_4=1$ and step profile of $J_2=0.03$ starting from initial
temperature $\beta_i=20$.  Similarly, Figure (\ref{fig:q4qq2bb20TEMP})
is for bump quench with $J_4=1$ and bump profile of $J_2=0.3$ from
$t=1$ to $t=10$ starting from initial temperature $\beta_i=20$.  In
all the other quenches, the stationary limit fits thermal ensemble
very well as in these two examples.

Since we observe thermalization, another observable of interest is
$G^>(t,t_2)$ where $t_2$ is fixed.  In the hydrodynamics
limit\cite{Bhattacharyya:2009uu} of large $t$, both the real and the
imaginary parts of the expectation value of this observable are again
exponential functions with both the exponents equal.  We will consider
the exponent of the imaginary part which we denote by $\gamma_{It}$.
This exponent is equal to the exponent of the retarded Green's
function in a thermal ensemble with temperature equal to the
temperature of the final thermalized limit of the quench process. We
will denote the exponent of the retarded Green's function by
$\gamma_{ret}$.
\begin{equation}
Im[G^>(t,t_2)]\xrightarrow{t\to \infty} a_2+b_2 e^{-\gamma_{It}t},
\qquad G^R(t,\beta_f)\xrightarrow{t\to \infty} a_3+b_3
e^{-\gamma_{ret}t}\ 
\end{equation}
At low temperature, $\gamma_{It}$ is proportional to the final temperature.
\begin{equation}
\gamma_{It}=\gamma_{ret} \sim \frac{\pi}{2\beta_f}\
\label{gammaIt}
\end{equation}

In a thermal ensemble, the retarded Green's function is a function of
the relative time difference.  In the conformal limit of SYK model,
the retarded Green's function in a thermal ensemble of inverse
temperature $\beta$ is
\begin{eqnarray}
  G^R(t_1,t_2)=&&-i2b\cos(\pi\Delta)\left(\frac{\pi}
                  {\sinh(2\pi(t_1-t_2)/\beta)}\right)^{\Delta}\,\theta(t_1-t_2)\nonumber\\
               &&\xrightarrow{(t_1-t_2)\to\infty}-i2b\cos(\pi\Delta)(2\pi)^{\Delta}
                  \,e^{-2\pi\Delta t/\beta}\,\theta(t_1-t_2)\
                  \label{eq:QQ-SYK_n3:2}
\end{eqnarray}
where $\Delta=1/q=1/4$ and $b=(4\pi J_4^2)^{-1/4}$.  In the conformal
limit, the exponent is 
\begin{equation}
\gamma_{conf}=\frac{2\pi\Delta}{\beta}=\frac{\pi}{2\beta}\ .
\label{gammaconf}
\end{equation}

\begin{figure}[ht]
\centering
\includegraphics[width=.5\linewidth]{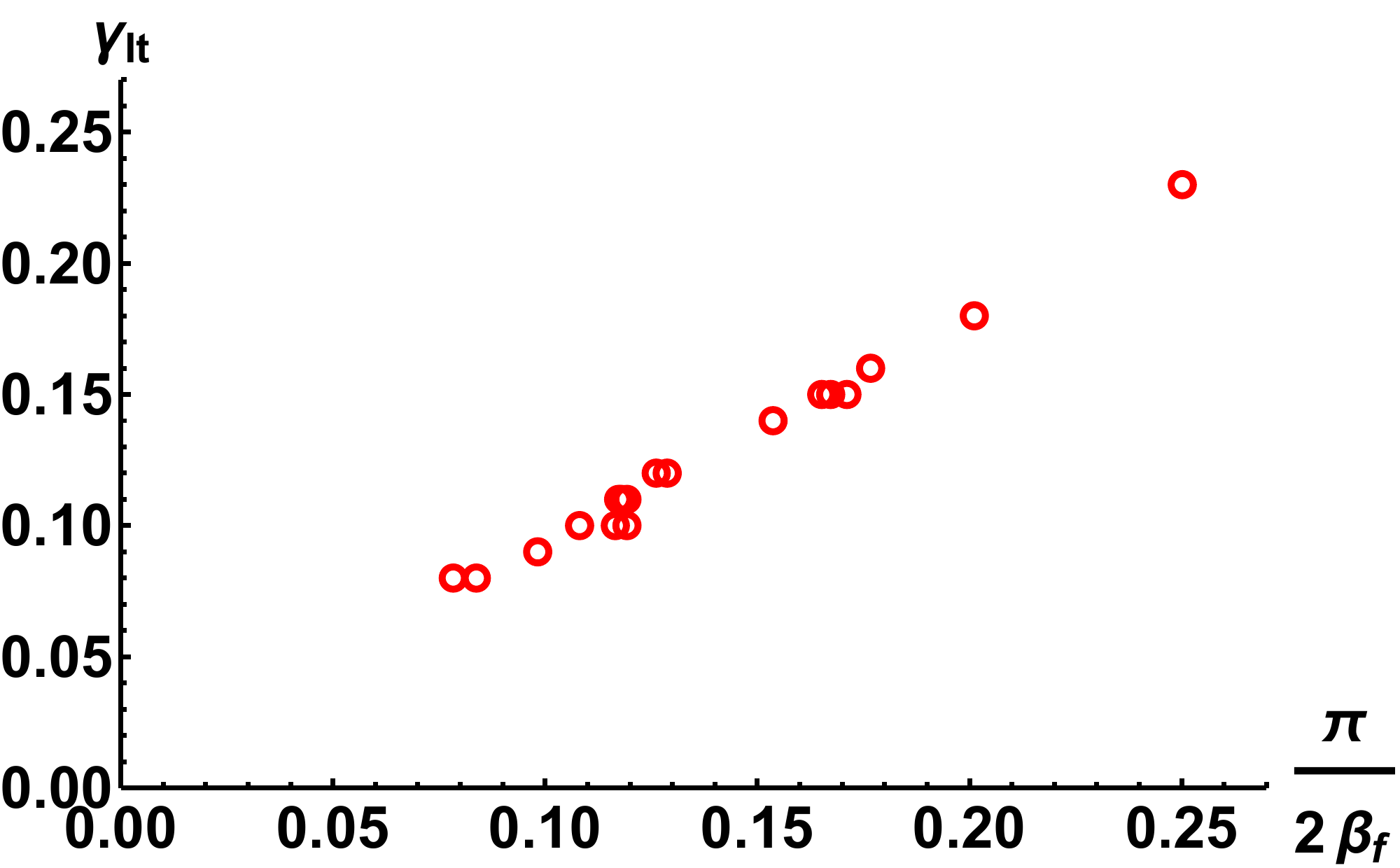}
\caption{The exponent $\gamma_{It}$ as a function of
  $\gamma_{conf}=\pi/(2\beta_f)$. $\gamma_{ret}$ is exactly equal to
  $\gamma_{It}$ as we can see from Table \ref{table:q4} so
  $\gamma_{ret}$ is not plotted here.}
\label{fig:JgammaIt}
\end{figure}

Figure (\ref{fig:JgammaIt}) is the plot of $\gamma_{It}$ and $\gamma_{conf}$.

At high temperatures, we find that the exponent of $G^R(t)$ gets
significant correction compared to its value at the conformal limit.
The corrected value of the exponent, which we have denoted by
$\gamma_{ret}$ above, is calculated by solving the SD equation
numerically.

Important numerical results for the step and bump quenches with $J_2$,
starting from different initial temperatures, are summarized in Table
\ref{table:q4}.
\begin{table}
  \caption{Numerical results for different quench protocols in $q=4$
    theory by changing $J_2$ coupling.  The following are absolute
    values after taking care of the time step $dt=0.05$. The value of
    $J_4$ is fixed during the entire quench process. The values of
    $J_2$ are the perturbations used to perform the different quench
    protocols.}
\centering 
\begin{tabular}{c c c c c c c c c}
\hline
\hline
  $J_4$ & Quench & $J_2$ & $\beta_i$ & $\beta_f$ & $\gamma_{Itt}$
  & $\gamma_{It}$ & $\gamma_{ret}$ & $\gamma_{conf}$\\ [0.5ex]
\hline
0.5 & Bump & 0.1 & 20 & 18.75 & 0.50 & 0.08 & 0.08 & 0.08 \\
0.5 & " & 0.3 & 20 & 13.17 & 0.53 & 0.10 & 0.10 & 0.12 \\
0.5 & Step & 0.05 & 20 & 13.48 & 0.53 & 0.10 & 0.10 & 0.12 \\
1.0 & Bump & 0.1 & 10 & 9.39 & 1.08 & 0.15 & 0.15 & 0.17 \\
1.0 & " & 0.1 & 20 & 13.17 & 1.06 & 0.11 & 0.11 & 0.12 \\
1.0 & " & 0.2 & 20 & 10.22 & 1.06 & 0.14 & 0.14 & 0.15 \\
1.0 & " & 0.3 & 20 & 7.81 & 1.12 & 0.18 & 0.18 & 0.20 \\
1.0 & " & 0.3 & 30 & 12.45 & 1.00 & 0.12 & 0.12 & 0.13 \\
1.0 & Step & 0.03 & 10 & 9.51 & 1.16 & 0.15 & 0.15 & 0.17 \\
1.0 & " & 0.03 & 20 & 14.53 & 1.16 & 0.10 & 0.10 & 0.11 \\
1.0 & " & 0.04 & 10 & 9.18 & 1.14 & 0.15 & 0.15 & 0.17 \\
1.0 & " & 0.04 & 20 & 13.32 & 1.15 & 0.11 & 0.11 & 0.12 \\
1.0 & " & 0.05 & 20 & 12.20 & 1.14 & 0.12 & 0.12 &  0.13 \\
1.0 & " & 0.05 & 30 & 13.39 & 1.18 & 0.11 & 0.11 &  0.12 \\
1.5 & Bump & 0.1 & 10 & 8.89 & 1.68 & 0.16 & 0.16 & 0.18 \\
1.5 & " & 0.1 & 20 & 15.99 & 1.54 & 0.09 & 0.09 & 0.10 \\
1.5 & " & 0.1 & 30 & 20.05 & 1.53 & 0.08 & 0.08 & 0.08 \\
1.5 & " & 0.3 & 10 & 5.31 & 1.73 & 0.26 & 0.26 & 0.30 \\
1.5 & " & 0.3 & 20 & 6.28 & 1.66 & 0.23 & 0.23 & 0.25 \\ [0.5ex]
\hline
\end{tabular}
\label{table:q4}
\end{table}
We also calculate the exponent $\gamma_{Itt}$ for $G^>(t,t-100)$,
$G^>(t-300,t)$, $G^>(t,t-300)$, $G^>(t-500,t)$ and $G^>(t,t-500)$.
The numerical values do not change significantly compared to the values
given in Table \ref{table:q4} for $G^>(t-100,t)$ hence, we can
conclude that $G^>(t-t_a,t)$ and $G^>(t,t-t_a)$ thermalize
exponentially with the same exponent for arbitrary $t_a$.

\begin{figure}[h]
\centering
\subcaptionbox{\label{fig:betaeff}}
  [.5\linewidth]{\includegraphics[width=.5\linewidth]{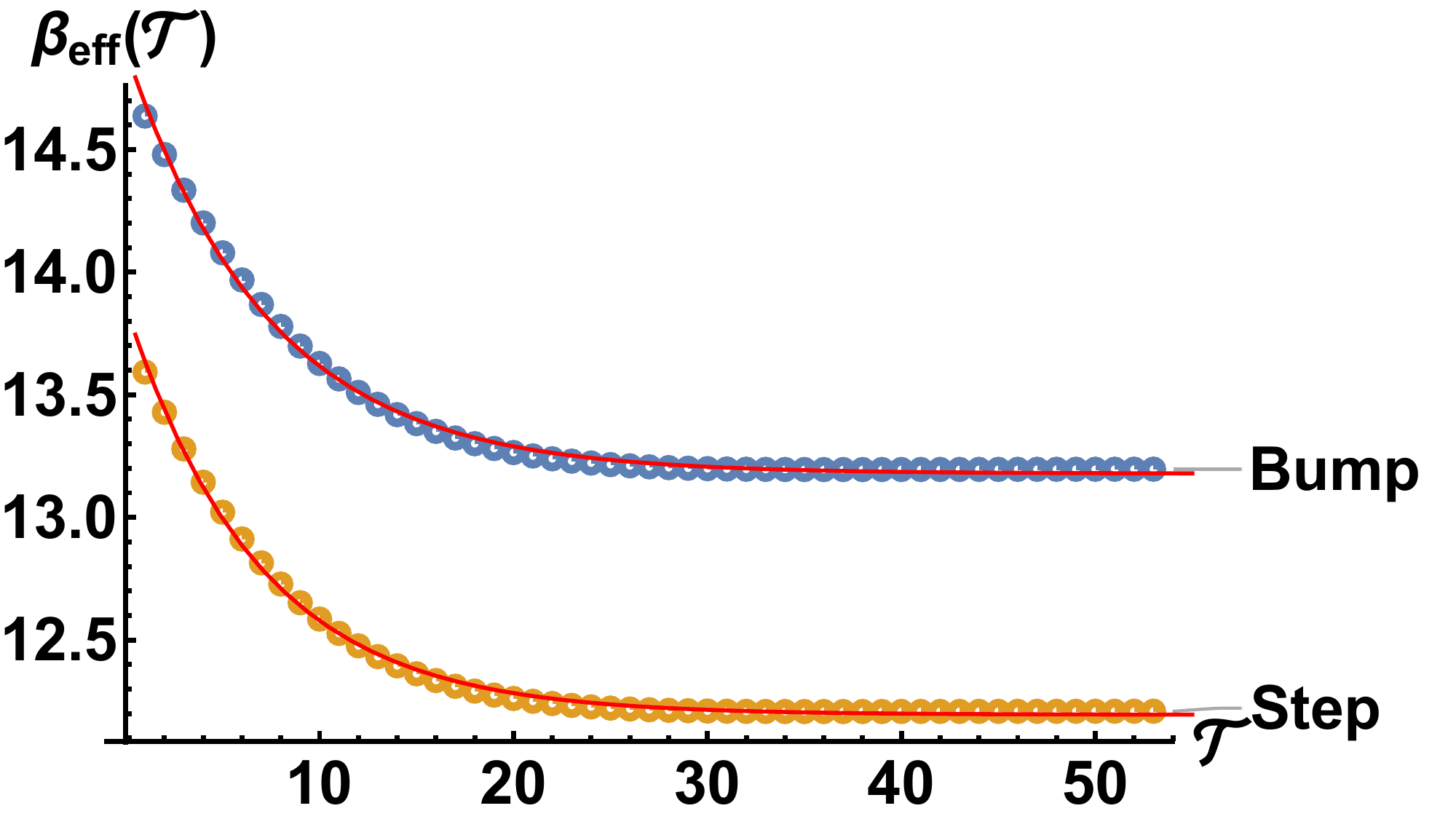}}
\subcaptionbox{\label{fig:invb_Gamma}}
  [.49\linewidth]{\includegraphics[width=.45\linewidth]{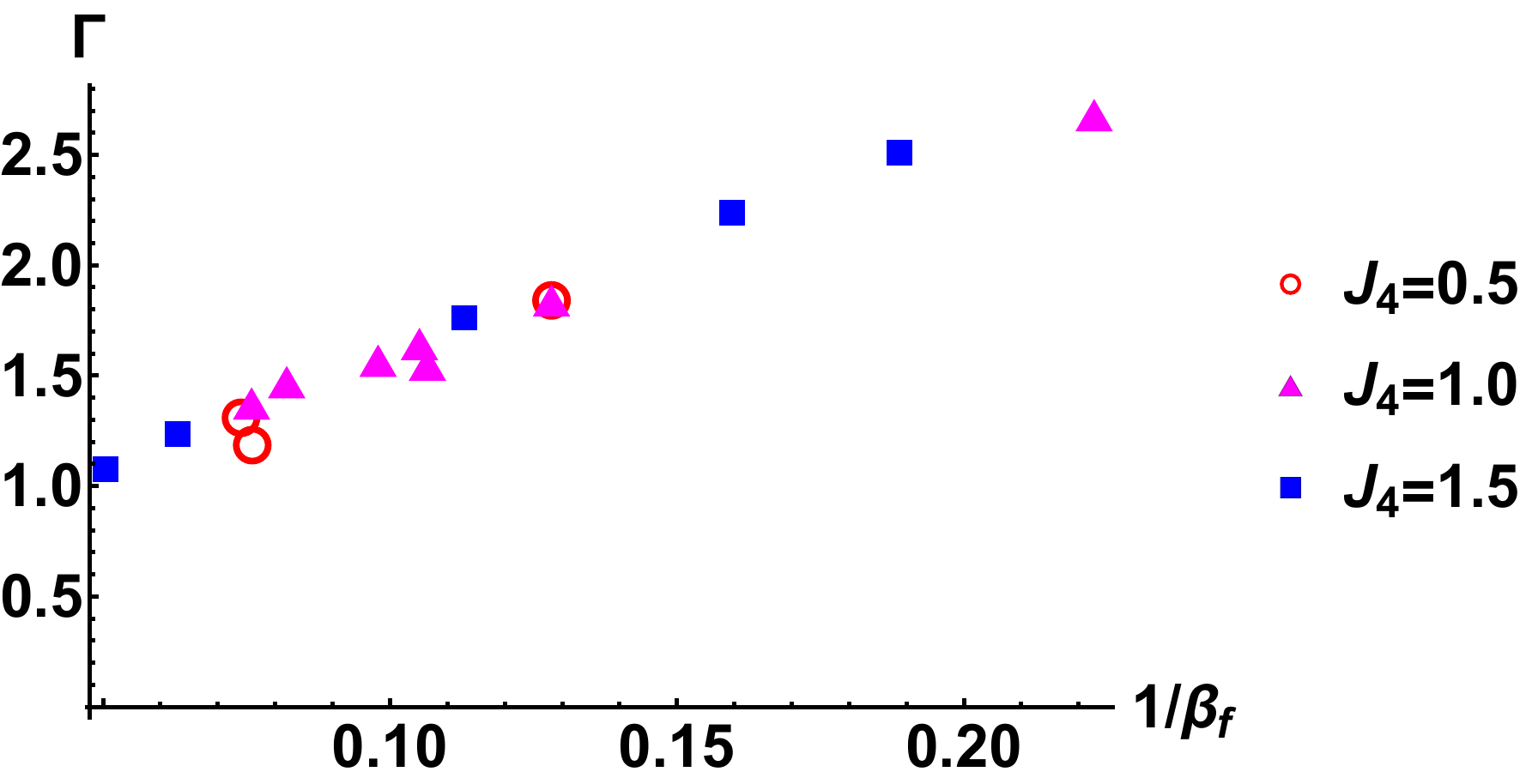}}
  \caption{(a) Examples of exponential fit of
    $\beta_{eff}(\mathcal{T})$ where $\mathcal{T}$ is in units of
    $2\times dt = 0.1$,  (b) The thermalization rate $\Gamma$ as a function of
    final temperature $1/\beta_f$ for different values of $J_4$ using step as well as bump quench protocols.}
\end{figure}

Let us now look at the thermalization rate of the effective
temperature in the case of step as well as bump quench\footnote{We
  thank the anonymous referee for drawing our attention to the computation
  of the thermalization rate.}.  Once the system thermalizes
it acquires the final equilibrium temperature, which we denote as
$1/\beta_f$.  Following \cite{Eberlein:2017wah}, we assume that the
relaxation behaviour of the effective temperature is given by
\begin{equation}
  \label{eq:QQ_SYK_rev2:2}
     \beta_{eff}(\mathcal{T}) = \beta_f + \alpha\exp{(-\Gamma \mathcal{T})}, \qquad \Gamma=\frac{C}{\beta_f}\ .
\end{equation}
The $\beta_{eff}$ settles down to the $\beta_f$ in the long time
limit.  Therefore one needs to determine $\alpha$ and $\Gamma$ in
\eqref{eq:QQ_SYK_rev2:2}.  To do that we make a change of variables
from $(t_1,t_2)$ to $\mathcal{T}=(t_1+t_2)$, and $t=t_1-t_2$ and
analyse the ratio of the Keldysh Green's function with the spectral
function for different values of $t$ but holding $\mathcal{T}$ fixed.
We repeat this for different values of $\mathcal{T}$.  It is
convenient to work in the frequency space, therefore we carry out a
partial Fourier transform with respect to $t$ and compute
\begin{equation}
  \label{eq:QQ_SYK_rev2:3}
  \frac{iG^K(\mathcal{T},\omega)}{A(\mathcal{T},\omega)}\ .
\end{equation}
We then calculate the effective temperature by taking small $\omega$
limit of the above quantity by fitting it to a hyperbolic tangent
function as given in \eqref{effbeta}.  The effective
temperature obtained in this procedure is then fitted to the
exponential ansatz given in \eqref{eq:QQ_SYK_rev2:2}.  We summarise
our results in the figure (\ref{fig:betaeff}) and
(\ref{fig:invb_Gamma}).
%


It is clear from Fig. \ref{fig:invb_Gamma} 
that like in the case of the step quench  as was observed in \cite{Eberlein:2017wah}, the thermalization rate
$\Gamma$ for the bump quench is also proportional to the final
temperature. It is also interesting to note that the constant $C$ is independent of the coupling $J_4$.

\section{Conclusion and Discussion}
\label{sec:condis}
We studied quench in the SYK model with different quench protocols.
While we have presented results for $q=2$ theory, and $q=4$ theory
with step and bump quench protocols, we have carried out this analysis
for $q=6$ as well as for $q=8$ models.  We find that the qualitative
features of the results are similar to the $q=4$ cases.

We observed that the $q=2$ theory does not thermalize for any of the
quench scenario we considered.  We considered quenching of $J_4$,
$J_6$, and $J_8$ using step and bump protocol. The initial states that
we considered are thermal states of inverse temperature
$\beta_i=10, 20$, and $30$ as well as the ground states. An
interesting aspect of all the quenches is that the greater Green's
function $G^>(t_1,t_2)$ equilibrates instantaneously as shown in
(\ref{q2_notherm}). Its expectation value freezes once both the time
arguments are outside the quench region.  Although in the final states
$G^>(t_1,t_2)$ equilibrates instantaneously, its equilibrium value is
not the same as the thermal ensemble expectation value.

The instanteneous equilibation or freezing that we observed is like a
glassy state.  It can be shown that if the final theory have both
$J_2$ and $J_4$ couplings, then two point functions always thermalize.
This is true even for arbitrarily small $J_4$ coupling in the large N
limit that we are considering.  We expect that this would change if we
consider effects subleading in N, where $J_2$ and $J_4$ couplings would
truly start competing \cite{PhysRevLett.120.241603}.

It would be interesting to identify the final state after each of
these quenches.  It is, however, beyond the scope of the present work
since we are working only with the equations of motion of the
$G^>(t_1,t_2)$ and solving them as an initial value problem.
  The $q=2$ theory is not chaotic and does not satisfy the ETH,
nevertheless thermalization in this theory is possible if the final
state were a KM state (\ref{kmstate}). This, for example, happens quite
often with step quenches in $1+1$ dimensional theories (even in
integrable theories) where the analog of KM states are the CC states
(\ref{ccstate}).


In $q=4$ theory, we find that thermalization happens in all the quench
scenario we considered.  We considered quenching with $J_2$, $J_6$,
and $J_8$ using step and bump protocols.  The initial states are
thermal states of inverse temperature $\beta_i=10, 20$, and $30$.  We
examined two kinds of greater Green's functions, $G^>(t-t_a,t)$ and
$G^>(t,t_b)$ as a function of time $t$ with fixed $t_a$ and $t_b$.

When both the time arguments $t-t_a$ and $t$ are outside the quench
region, both the real and imaginary parts of $G^>(t-t_a,t)$ are
exponential functions with the same exponent. This exponent
$\gamma_{Itt}$ is equal to the value of coupling $J_4$ of the system.
It would be interesting to compare this result with the bulk
calculation.  The scalar Feynman propagator is known to thermalize
instantaneously in the AdS$_2$-Vaidya
spacetime\cite{Ebrahim:2010ra, Keranen:2014lna}.
This instantaneous thermalization is seen only for AdS$_2$ and is not
observed, say, in AdS$_3$.  There is no analogous calculation for
fermions is available in the literature.

\begin{figure}[h]
\begin{center}
\includegraphics[width=0.4\textwidth]{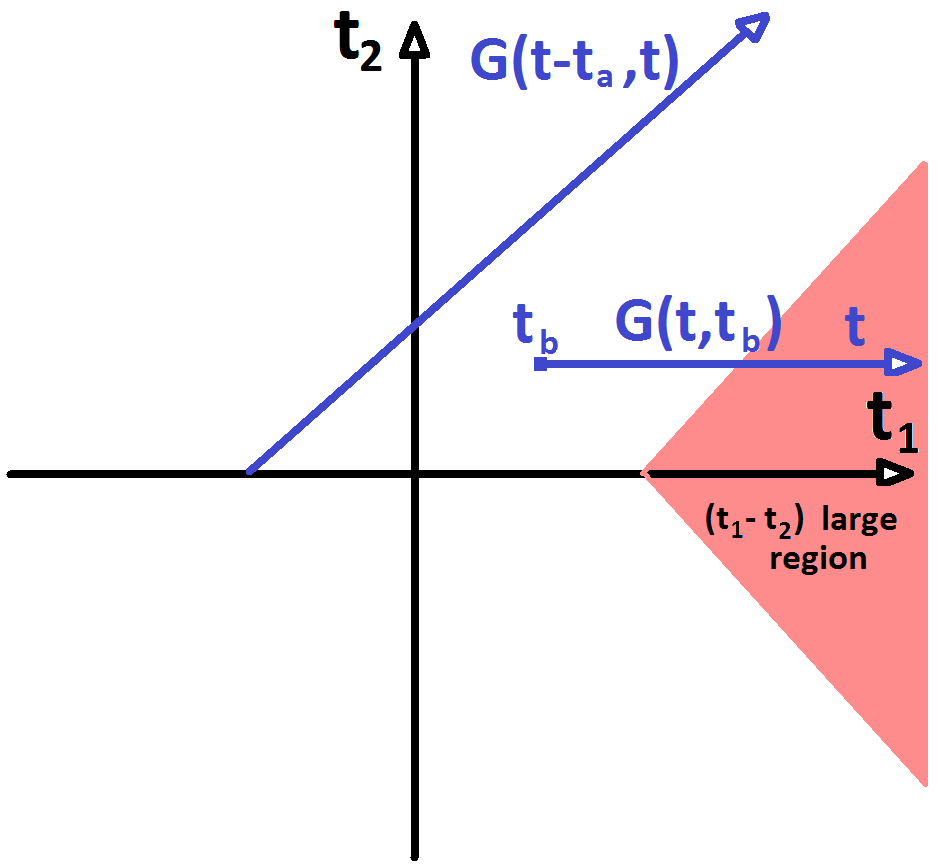}
\caption{\small The large $t$ limit of greater Green's function
  $G^>(t,t_b)$ is calculated in the large $(t_1-t_2)$ region of the
  retarded Green's function $G^R(t_1,t_2)$. Moreover, in this region,
  the system has more or less thermalized. Hence,
  $\gamma_{It}=\gamma_{ret}$.}
\label{fig:grid_KBgrt}
\end{center}
\end{figure}

The long time limit of both the real and imaginary parts of
$G^>(t,t_b)$ are exponential functions with the same exponent. This
exponent $\gamma_{It}$ is equal to the exponent $\gamma_{ret}$ of the
retarded Green's function $G^R(t_1,t_2)$ in a thermal ensemble
(\ref{eq:QQ-SYK_n3:2}) with temperature equal to the final temperature
of the quench process. This is obvious at least for the imaginary part
of the $G^>(t,t_b)$ since the system thermalizes.  $G^R(t_1,t_2)$ is a
simple multiple of the imaginary part of $G^>(t_1,t_2)$. As one can
see in Figure \ref{fig:grid_KBgrt}, the long time limit of
$G^>(t,t_b)$ is calculated in a subset of the large $(t_1-t_2)$ of
$G^R(t_1,t_2)$.

We also studied the thermalization rate for both step and bump
protocols.
In both cases we find that the thermalization rate
is proportional to the final temperature $1/\beta_f$ and the proportionality constant $C$ is independent of the coupling constant of the final theory.

One clear and important observation that we can make is that
the thermalization in $q=4$ theory is not because the final state is a
KM state. If the final state had been a KM state, $G^>(t-t_a,t)$ would
have thermalized instantaneously once both its time arguments are
outside the quench region.\footnote{Although quenches in $q=4$ theory
  start from thermal states, as we have noted above the results should
  be qualitatively similar and quantitatively close to quenches
  starting from ground states.} This is because the `diagonal'
two-point function that we are considering are already thermalized in
a KM state.

\subsection{How to prepare KM states}
\label{kmprep}
The KM state, in principle, can be prepared by performing a
{\emph{sudden}} quantum quench starting from the ground state using
the extra term 
\begin{equation}
H_{\mu}(t)=i\mu(1- \Theta(t)) \sum_{k=1}^{N/2} s_k \psi_k(t)\psi_{k+1}(t)
\label{hkm}
\end{equation}
where $s_k$'s specifies the particular $|B_s\rangle$ defined in
(\ref{bs_state}).  This new term has been used in a different but
related context in \cite{Kourkoulou:2017zaj}.  The argument behind
this assertion is similar to the argument provided in
\cite{Maldacena:2018lmt} for the preparation of thermofield double
state by performing a sudden quench.  We will consider small $\mu$
limit.  The full Hamiltonian before the quench at $t=0$ is
\begin{equation}
H+H_{\mu}=(i)^{q/2}\sum_{1\leq i_1< i_2<...< i_q\leq
  N}J_{i_1,i_2,..,i_q}\psi_{i_1}\psi_{i_2}....\psi_{i_q}+i\mu
\sum_{k=1}^{N/2}s_k \psi_k\psi_{k+1}\
\end{equation}
The ground state of the above Hamiltonian is the state which minimizes
the second term.  But minimizing the second term corresponds to strong
positive or negative correlation of $\psi_k$ and $\psi_{k+1}$
depending on the value of $s_k$.  Strong correlation of $\psi_k$ and
$\psi_{k+1}$ is the basis of the definition of $|B_s\rangle$ in
(\ref{bs_state}).  In hindsight, it is in some sense obvious why the
KM states were not obtained from the step quench using the disordered
couplings like $j_{2,ij}$ or $J_2$.  This is because not just two
fermions, but all the fermions were randomly and strongly correlated
in the ground states of the initial Hamiltonians.

\subsection{Ergodicity versus Mixing}
In this work, we don't consider long time averaging. $q=4$ theory
satisfies eigenstate thermalization hypothesis(ETH). But
thermalization from ETH crucially requires long time
averaging. Thermalization without long time averaging has been
observed in many other works but in most of the cases it is because
the final state turns out to be a very particular state like CC
states. So, in this sense, the thermalization that we observed is much
more robust than what one expects from ETH. Thermalization with ETH
follows from quantum ergodicity. But what we observe is more akin to a
quantum version of mixing.

\begin{figure}[h]
\centering
\subcaptionbox{Ergodicity \label{fig:ergo}}
  [.4\linewidth]{\includegraphics[width=.4\linewidth]{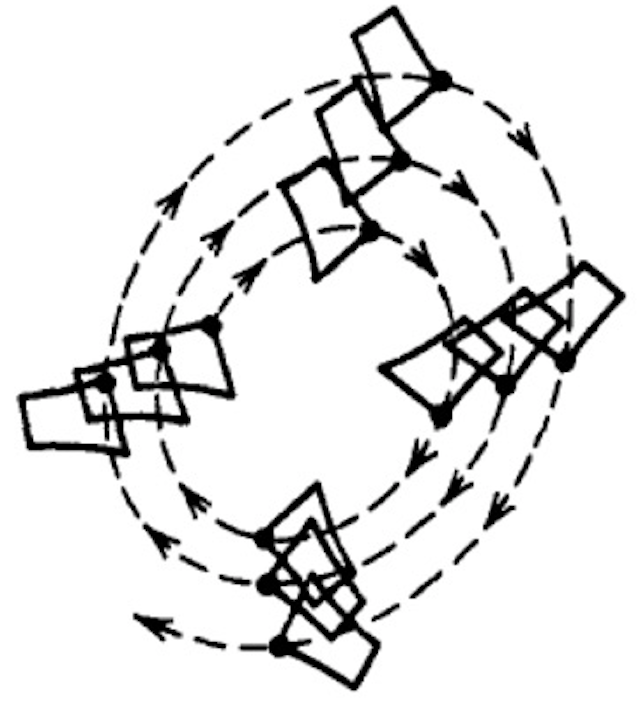}}
\subcaptionbox{Mixing \label{fig:mixing}}
  [.4\linewidth]{\includegraphics[width=.4\linewidth]{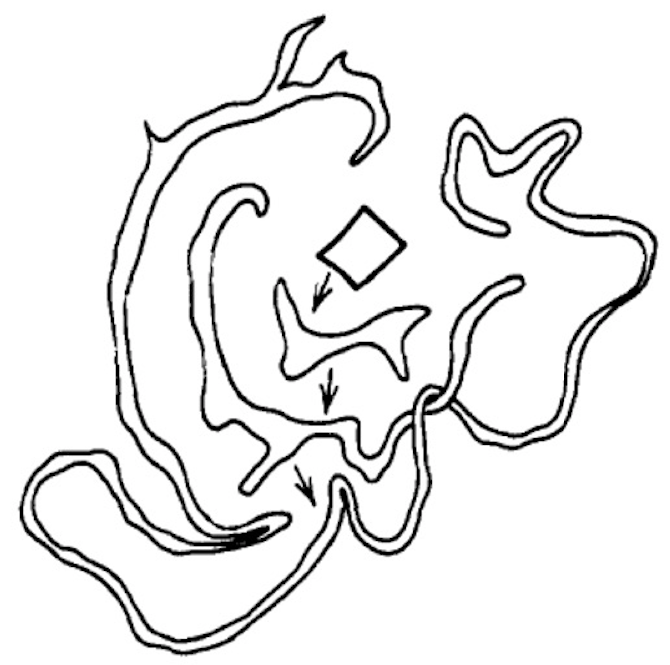}}
  \caption{(a) Ergodicity: the shape of the initial sample only
    changes slightly but sweeps out the entire allowed region under
    time evolution, (b) Mixing: the initial sample spreads out and
    reaches infinitesimally close to all the points in the allowed
    region of the phase space. Figure adopted from
    \cite{zbMATH01736203,ziraldo2013thermalization}.}
\end{figure}

In classical theories, mixing is a much stronger phenomenon compare to
ergodicity. Figure (\ref{fig:ergo}) shows ergodic evolution in the
classical phase space. The initial state is described by an ensemble
concentrated in the deformed rectangle in the phase space. The volume
is conserved under time evolution due to the Liouville theorem for a
closed system, but the shape can change. For ergodic systems, the
shape of the initial sample hardly changes but it sweeps out the
entire allowed space under time evolution. So, a long time averaging
gives the expectation value in the micro-canonical ensemble.  In
mixing, as shown in Figure (\ref{fig:mixing}), the initial sample
spreads out and reaches infinitesimally close to all the points in the
allowed region of the phase space. So, without time averaging, mixing
gives the expectation value in the microcanonical ensemble.

Using this classical analogy, we believe that even in quantum systems,
chaos is a much stronger condition for thermalization than the
eigenstate thermalization hypothesis(ETH).  Our results on
thermalization in the quenched SYK model seem to suggest that
quench without long time average is a quantum analog of mixing.  It
would be interesting to make this more concrete.  We hope to return to
this soon.


\section*{Acknowledgement}
We would like to thank Sumit Ranjan Das, Juan Maldacena, Gautam
Mandal, Ganapathy Murthy, Sumathi Rao, and Ashoke Sen for discussion.
We would also like to thank Julia Steinberg for sharing their main
program.  We would like to thank the anonymous referee for
illuminating remarks which helped improve our results.  This research
was supported in part by the International Centre for Theoretical
Sciences (ICTS) during a visit for participating in the program -
AdS/CFT at 20 and Beyond (Code: ICTS/adscft20/05).

\bibliography{sykquench} 
\bibliographystyle{JHEP}

\end{document}